\documentclass[12pt]{article}
\usepackage{amsmath,amssymb}
\newcommand{\be}{\begin{equation}}
\newcommand{\ee}{\end{equation}}
\newcommand{\bea}{\begin{eqnarray}}
\newcommand{\eea}{\end{eqnarray}}
\newcommand{\nn}{\nonumber \\}
\newcommand{\p}[1]{(\ref{#1})}
\newcommand{\lb}{\label}

\newcommand{\und}{\underline}

\begin{document}
\begin{titlepage}
\begin{flushright}
JINR E2-2011-69
\end{flushright}
\vskip 0.6truecm

\begin{center}
{\Large\bf ${\cal N}{=}4$ mechanics of general (4, 4, 0)
multiplets} \vspace{1.5cm}

{\large\bf F. Delduc$\,{}^a$, E. Ivanov$\,{}^b$,}\\
\vspace{1cm}

{\it a)Laboratoire de Physique, CNRS et Universit\'e de Lyon, ENS Lyon,}\\
{\it 46, all\'ee d'Italie, 69364 LYON Cedex 07, France}\\
{\tt francois.delduc@ens-lyon.fr}
\vspace{0.3cm}

{\it b)Bogoliubov  Laboratory of Theoretical Physics, JINR,}\\
{\it 141980 Dubna, Moscow region, Russia} \\
{\tt eivanov@theor.jinr.ru}\\

\end{center}
\vspace{0.2cm}
\vskip 0.6truecm  \nopagebreak

\begin{abstract}
\noindent We construct the manifestly ${\cal N}=4$ supersymmetric off-shell superfield ``master'' action for any number $n$ of the ${\cal N}=4$
supermultiplets $({\bf 4, 4, 0})$ described by harmonic analytic superfields $q^{+a}(\zeta, u), a= 1, \ldots 2n\,,$ subjected
to the most general harmonic constraints. The action consists of the sigma-model and Wess-Zumino parts. We present
the general expressions for the target space metric, torsion and background gauge fields. The generic target space geometry
is shown to be weak HKT (hyper-K\"ahler with torsion), with the strong HKT and HK ones as particular cases. The background gauge fields
obey the self-duality condition.
Our formulation suggests that the weak HKT geometry is fully specified by the two primary potentials:
an unconstrained scalar potential ${\cal L}(q^+, q^-, u)|_{\theta = 0}$ which is the $\theta = 0$ projection
of the superfield sigma-model Lagrangian, and a  charge 3 harmonic analytic potential ${\cal L}^{+ 3a}(q^+, u)|_{\theta = 0}$ coming from
the harmonic constraint on $q^{+ a}\,$. The reductions to the strong HKT and HK geometries amount to simple restrictions on the
underlying potentials.
We also show, using the ${\cal N}=2$ superfield approach, that the most general bosonic target geometry of the
${\cal N}=4, d=1$ sigma models, of which the weak HKT geometry is a particular case, naturally comes out after adding the mirror
$({\bf 4, 4, 0})$ multiplets with different transformation laws under ${\cal N}=4$ supersymmetry and $SO(4)$ $R$ symmetry. Thus the minimal
dimension of the target spaces exhibiting such a ``weakest'' geometry is 8, which corresponds to a pair of the mutually mirror
$({\bf 4, 4, 0})$ multiplets.
\end{abstract}
\vspace{0.7cm}

\newpage

\end{titlepage}

\section{Introduction}
It is widely known that the superfield formulations of supersymmetric sigma models in diverse dimensions, with all involved supersymmetries
being manifest and off-shell, reveal the underlying target bosonic geometries of these models in the most clear way. Such formulations
directly lead to the basic unconstrained potentials of the target geometries, and this is one of their basic advantages.
In most of cases, the potential appears as the generic superfield Lagrangian of the given supersymmetric sigma model. For instance,
the most general sigma-model type Lagrangian of ${\cal N}=1, d=4$ chiral superfields (or of their ${\cal N}=2, d=3$, ${\cal N}=(2,2), d=2$ and
${\cal N}=4, d=1$ reductions) is just the K\"ahler potential of the underlying target K\"ahler geometry \cite{Zum}. Analogously,
the most general Lagrangian of the analytic hypermultiplet superfields $q^{+A}(\zeta, u)$ in the ${\cal N}=2, d=4$ harmonic superspace
formulation \cite{HSS,HSS1} is an unconstrained analytic function ${\cal L}^{+4}(q^{+ A},u^\pm_k)$ which encodes all the information
about the underlying target hyper-K\"ahler (HK)
geometry \cite{AGF}. For this reason and by analogy with the ${\cal N}=1, d=4$ case, this object was called the
``hyper-K\"ahler potential'' in \cite{hkHSS}. The
analogous harmonic analytic superfield Lagrangian, extended to couplings to the most general Poincar\'e ${\cal N}=2$ supergravity \cite{sgN2},
is the fundamental object of the corresponding quaternion-K\"ahler (QK) target geometry \cite{BaWi},
whence the name ``quaternion-K\"ahler potential'' for it \cite{qkHSS}.
For the HK and QK geometries, the relevant potentials were firstly introduced, when constructing the general off-shell superfield actions
for the hypermultiplets in the harmonic superspace (both in the flat case and in the ${\cal N}=2$ supergravity background).
Later on, these potentials were recovered \cite{geom1}, \cite{qkHSS} as the underlying objects of the corresponding geometries  by constructing
the appropriate harmonic extensions of HK and QK manifolds and then solving the standard constraints of these geometries
in such extended  spaces.

One more important example of how manifestly supersymmetric formulations help to reveal the unconstrained potentials
of the target geometries is offered by ${\cal N}= (4,0), d=2$ supersymmetric hypermultiplet  sigma models.
It was known that the relevant target geometry is the strong
version of the so-called hyper-K\"ahler geometry with torsion (strong HKT)\cite{hkt} \footnote{This sort of target space geometries
was revealed in the context of supersymmetric sigma models much earlier than the nomenclature HKT was suggested for them
(see e.g. \cite{402,400,401,404,chs,DV,BV,popad2}). In particular, they are target geometries of
${\cal N}=(4,4), d=2$ supersymmetric group manifold WZNW sigma models \cite{400,404}.}.
The corresponding action was constructed in \cite{dks}
in the harmonic superspace. It was found to be fully specified by the analytic unconstrained potential ${\cal L}^{+ 3 a}(q^+, u)$, $a=1, \ldots 2n$
and $n$ being the number of the hypermultiplets involved. As was also shown in \cite{dks}, the same object naturally appears within
the pure geometric setting: it solves the general
constraints of the  ${\cal N}= (4,0), d=2$  sigma model geometry \footnote{Strictly speaking, there appears  one more potential,
but it can be gauged into its flat value by the proper target space gauge transformations.}.

In the present paper we apply a similar strategy to the case of the most general ${\cal N}=4$ supersymmetric mechanics based on the
off-shell multiplets ${\bf (4, 4, 0)}$ (the numerals here stand for the numbers of the physical bosonic,
physical fermionic and auxiliary fields). This multiplet is an important ingredient of the ${\cal N}=4$ supersymmetric quantum mechanics (SQM)
model-building. It can be treated as a ``root'' ${\cal N}=4, d=1$ multiplet \cite{Root,root}, in the sense that SQM models associated with other
off-shell ${\cal N}=4, d=1$ multiplets (e.g., with multiplets ${\bf (3, 4, 1)}$ or ${\bf (2, 4, 2)}$) can be obtained from the ${\bf (4, 4, 0)}$ SQM
models via the so-called ``automorphic duality'' \cite{GR0}, in its linear  \cite{GR0,PT} or nonlinear \cite{IL,IKL1} versions.
As shown by us in \cite{gpr}, the manifestly supersymmetric superfield formulation of this sort of duality amounts to the procedure of gauging
various isometries of the ${\bf (4, 4, 0)}$ SQM models by some non-propagating (``topological'') ${\cal N}=4$ gauge superfields.

The natural description of the ${\bf (4, 4, 0)}$ multiplets is achieved in the ${\cal N}=4, d=1$ HSS \cite{IL} where they are represented by
the analytic superfields $q^{+ a}(\zeta, u)$ subjected to some harmonic constraints, which can be linear or non-linear. The general sigma-model type
actions  based on the linear ${\bf (4, 4, 0)}$ multiplets were constructed in \cite{IL}. The corresponding target-space metric
is conformally flat in the case of one such multiplet, i.e. for 4-dimensional target manifolds, and presents some natural generalization
of the conformally flat metric for the general $4n$-dimensional targets. Non-trivial target metrics, in particular,
the hyper-K\"ahler ones, can be gained, if alternatively dealing with the nonlinear ${\bf (4, 4, 0)}$ multiplets. Some special class of nonlinear
${\bf (4, 4, 0)}$ multiplets giving rise to SQM models with the 4-dimensional target hyper-K\"ahler metrics
described by the renowned Gibbons-Hawking ansatz \cite{GH} was considered in \cite{gpr} (in the component approach, these models
were also constructed in \cite{ksch}).

The basic goals of the present paper are, first, to present the most general nonlinear ${\bf (4,4,0)}$ multiplets in the HSS approach,
second, to construct the most general sigma-model and Wess-Zumino (WZ) type superfield actions for such multiplets and, third, to reveal the
relevant target space geometry.

We show that the set of $n$ most general ${\bf (4,4,0)}$ multiplets is described
by the analytic superfields $q^{+ a}\,$, $a=1,\ldots , 2n\,$, subjected to the nonlinear harmonic constraint
\footnote{Our notations are the same as in our previous papers, e.g. in \cite{gpr} (see also Sect. 3).}
\be
D^{++} q^{+ a} = {\cal L}^{+ 3 a}(q^{+b}, u^\pm_i)\,, \lb{1}
\ee
with ${\cal L}^{+ 3 a}$ being an arbitrary function of $q^{+ a}$ and explicit harmonics.

The most general sigma-model type superfield action
of these $q^{+ a}$ multiplets is given by the integral over the whole ${\cal N}=4$ HSS
\be
S \sim \int du dt d^4\theta \,{\cal L}(q^{+ a}, q^{- b}, u^\pm_i)\,, \quad q^{- a} = D^{--}q^{+ a}\,. \lb{2}
\ee

The corresponding target space geometry is the general weak HKT geometry (see, e.g., \cite{CoPa,GiPaSt,Hu} for the precise
definitions \footnote{Discussion of HKT manifolds from the pure mathematical point of view can be found, e.g., in a recent preprint \cite{math}
(and refs. therein).}).
The HSS superfield formulation suggests that this geometry is specified by two general potentials: the analytic potential
${\cal L}^{+ 3 a}\vert $ (hereafter, $\vert$ denotes the restriction to the $\theta$ independent parts), and the scalar non-analytic
potential
\be
{\cal L}(q^{+ a}, q^{- b}, u^\pm_i)\vert\,. \lb{3}
\ee

The most general coupling to an external (abelian) gauge field is given by the superfield WZ term
\be
S_{WZ} \sim \int  du d \zeta^{(-2)} \,{\cal L}^{+ 2}(q^{+ a}, u^\pm_i)\,. \lb{4}
\ee

We also argue that the most general ${\cal N}=4, d=1$ sigma model geometry (which is even less restrictive than the weak HKT one \cite{GiPaSt,Hu})
corresponds to adding another, ``mirror'' (or ``twisted'') sort of the ${\bf (4,4,0)}$ multiplets \cite{BIKL,ivnie}.
They differ from those described by the superfields $q^{+ a}$ in that the roles of two independent $SU(2)$ automorphism
groups of the ${\cal N}=4, d=1$
Poincar\'e superalgebra are interchanged for them. We were not able to perform this more general analysis in the ${\cal N}=4$ superspace framework
\footnote{Such a study seems to require the more complicated bi-harmonic ${\cal N}=4, d=1$ superspace framework \cite{ivnie}.}
and did this, using the alternative ${\cal N}=2$ superfield formulation which extends the one given in \cite{Hu}.

The paper is organized as follows. In Sect. 2, as a warm-up, we recall basic facts about the ${\cal N}=1$ and ${\cal N}=2$
superfield formulations of sigma models in one dimension. The new result is the derivation of the potentials specifying the most general
${\cal N}=2, d=1$ superfield sigma model action \cite{Hu} directly from the constraints defining the relevant
target space geometry \cite{CoPa,GiPaSt,Hu},
like this has been done in \cite{geom1,qkHSS,dks} for other cases (already mentioned above). In Sect. 3 we recall the basics of
the ${\cal N}=4, d=1$ HSS approach and how the linear ${\bf (4,4,0)}$ multiplet is described within it. Sections 4 and 5 contain our basic results.
There we define the most general nonlinear ${\bf (4,4,0)}$ multiplet, construct its most general superfield sigma-model type and WZ
type actions and recover the geometry hidden in the sigma model action. It turns out to be general weak HKT. We also specify
the conditions under which it is reduced to the strong HKT
and general HK geometries. The background gauge field entering the WZ type action satisfies the self-duality condition, whatever the
target space geometry is. In Sect. 6 we collect some particular cases of interest. In Sect. 7 we come back
to the ${\cal N}=2$ superfield formulation in order to exhibit the conditions imposed on the most general ${\cal N}=2$ superfield
sigma-model action by the requirement that it possesses
an extra ${\cal N}=2$ supersymmetry which builds up the manifest ${\cal N}=2$ supersymmetry to the ${\cal N}=4$ one.
This analysis first repeats what has already been done in \cite{Hu}, but we do one step further by
including into the consideration, in the ${\cal N}=2$ superfield language, simultaneously two different sorts of the
${\bf (4,4,0)}$ multiplets. We show that in such an extended system (with at least 8-dimensional target space)
the requirement of hidden ${\cal N}=4$ supersymmetry gives rise just to the conditions of the most general ${\cal N}=4, d=1$ sigma model
target geometry, which are weaker than those of the weak HKT one.

\setcounter{equation}{0}

\section{${\cal N}=1$ and  ${\cal N}=2$ one-dimensional supersymmetry}
Before turning to the ${\cal N}=4$ case, let us remind the reader of some results that may be found in \cite{CoPa, GiPaSt,Hu}
where a generic ${\cal N}=2$ supersymmetric model in one dimension is studied. We start by writing down a generic model
with ${\cal N}=1$ supersymmetry.
One may use a superspace with coordinates $(t,\theta)$. The superfields $X^{i}(t,\theta)=x^{i}(t)+\theta\lambda^{i}(t)$ considered
here are bosonic. The variables $x^{i}$ may be seen as local coordinates on some manifold. The supersymmetric covariant derivative $D$ reads
\begin{equation}
D=\frac{\partial}{\partial \theta}+i\theta\partial_{t},\quad D^{2}=i\partial_{t},\quad\partial_{t}=\frac{\partial}{\partial t}.
\label{symalg}
\end{equation}
A general ${\cal N}=1$ supersymmetric action then reads
\begin{equation}
S=-\frac{1}{2}\int dtd\theta \,[ ig_{ij}(X)D X^{i}\partial_{t}X^{j}+\frac{1}{3!}c_{ijk}(X)
DX^{i}DX^{j}DX^{k}]\,,
\label{n1act}
\end{equation}
where the tensor $g_{ij}$ may be chosen symmetric\begin{footnote}{Using the algebra (\ref{symalg}), a term containing an antisymmetric
tensor $g_{ij}$ may, using integration by parts, be recast in the form of the second term on the right-hand
side of (\ref{n1act}).}\end{footnote} and will be interpreted as a metric on the target manifold. The 3-tensor $c_{ijk}$
is fully antisymmetric and will receive the interpretation of a torsion on the manifold. As opposed to models
in two dimensions, the torsion three-form $c=c_{ijk}(x)dx^{i}\wedge dx^{j}\wedge dx^{k}$ is not closed in one dimension.
In terms of the component fields $x^{i}(t)=X^{i}(t)\vert_{\theta=0}$, $\lambda^{i}(t)=DX^{i}(t)\vert_{\theta=0}$,
we rewrite the action (\ref{n1act}) as
\begin{equation}
S=\frac{1}{2}\int dt\,[g_{ij}(x)\partial_{t}x^{i}\partial_{t}x^{j}+ig_{ij}(x)\lambda^{i}\nabla\lambda^j
+\frac{1}{3!}c_{ijk,l}\lambda^{i}\lambda^{j}\lambda^{k}\lambda^{l}]\,,\quad f(x)_{,i}=\frac{\partial f(x)}{\partial x^{i}}\,.
\label{n1actc}
\end{equation}
The covariant derivative $\nabla$ is defined as :
\begin{equation}
\nabla\lambda^j=\partial_{t}\lambda^{j}+\Gamma^{j}_{kl}\partial_{t}x^{k}\lambda^{l},\quad
\Gamma^{j}_{kl}=\gamma^{j}_{kl}(g)+\frac{1}{2}g^{jm}c_{mkl},
\end{equation}
where $g^{ij}$ is the inverse of the metric $g_{ij}$, and $\gamma^{j}_{kl}(g)$ is the Christoffel symbol associated with the metric $g$.

\subsection{Geometrical constraints}
We shall now restrict the geometry of the manifold  by the requirement of extended supersymmetry. In particular,
it was shown in \cite{CoPa, GiPaSt,Hu} that the ${\cal N}=2$ supersymmetry algebra requires the existence
of a complex structure $I^{i}_{j}(x)$, $I^{i}_{j}I^{j}_{k}=-\delta^{i}_{k}$, with vanishing Niejenhuis tensor
$$I_{i}^{l}I_{[j,l]}^k-I_{j}^{l}I_{[i,l]}^k=0\,.$$
Invariance of the action (\ref{n1act}) under ${\cal N}=2$ supersymmetry requires that the metric is hermitian, $I_{k}^{i}g_{ij}I_{l}^{j}=g_{kl}$,
and that a symmetrized covariant derivative of the complex structure vanishes
\begin{equation}
\nabla_{(i}I^{j}_{k)}=0,\quad \nabla_{i}V^j(x)=\frac{\partial V^{j}}{\partial x^{i}}+\Gamma^{j}_{ik}V^{k}.
\label{symcov}
\end{equation}
Moreover, the torsion 3-form $c$ should satisfy the constraints
\begin{equation}
\iota_{I}dc-\frac{2}{3}d\iota_{I}c=0, \label{diffi}
\end{equation}
where $\iota_{I}$ is the inner derivation with respect to the complex structure $I$ defined, for any $p$-form $\omega^{(p)}$, by
\begin{equation}
\iota_{I}\omega^{(p)}(V_{1},\cdots,V_{p})=\sum_{q=1}^p\omega^{(p)}(V_{1},\cdots,IV_{q},\cdots,V_{p}).
\end{equation}

It turns out that these constraints on the geometry may be easily solved. First, the integrability of the complex
structure means that one can choose complex coordinates $z^\alpha$, $\bar z^{\bar\alpha}$ such that the complex structure takes a simple form
$$I_{\alpha}^{\beta}=i\delta_{\alpha}^{\beta},\, I_{\bar\alpha}^{\bar\beta}=-i\delta_{\bar\alpha}^{\bar\beta},\,
I_{\bar\alpha}^{\beta}=I_{\alpha}^{\bar\beta}=0.$$
The hermiticity of the metric then means that its only non-zero components are the mixed
ones $g_{\alpha\bar\beta}=g_{\bar\beta\alpha}$. The constraints (\ref{symcov}) then lead to constraints on the connection
\begin{equation}
\Gamma^{\bar\gamma}_{(\alpha\beta)}=\Gamma^{\gamma}_{(\bar\alpha\bar\beta)}=0,\,
\Gamma^{\gamma}_{\alpha\bar\beta}=\Gamma^{\bar\gamma}_{\bar\alpha\beta}=0.
\end{equation}
The first two of these equations are automatically satisfied, thanks to the hermiticity of the metric.
The last two allow to determine some components of the torsion tensor in terms of the metric
\begin{equation}
c_{\alpha\bar\beta\bar\gamma}=g_{\alpha\bar\beta,\bar\gamma}-g_{\alpha\bar\gamma,\bar\beta},\,
c_{\alpha\beta\bar\gamma}=g_{\alpha\bar\gamma,\beta}-g_{\beta\bar\gamma,\alpha}.\label{torco}
\end{equation}
We introduce the 2-form
\begin{equation}
\Omega=\frac{i}{2}g_{ij}I^{j}_{k}dx^{i}\wedge dx^{k}=g_{\alpha\bar\beta}dz^\alpha\wedge d\bar z^{\bar\beta},
\label{metfo}
\end{equation}
and separate the exterior derivative into holomorphic and anti-holomorphic parts
\begin{equation}
d=\partial+\bar\partial,\,\partial=dz^{\alpha}\frac{\partial}{\partial z^\alpha},\,
\bar\partial=d\bar z^{\bar\alpha}\frac{\partial}{\partial\bar z^{\bar \alpha}}.
\end{equation}
We also separate the torsion 3-form according to the number of holomorphic $dz^\alpha$
and anti-holomorphic $d\bar z^{\bar\alpha}$ differentials
\begin{eqnarray}&
c =c^{(3,0)}+c^{(2,1)}+c^{(1,2)}+c^{(0,3)},&\nonumber\\
&c^{(3,0)}=c_{\alpha\beta\gamma}dz^\alpha\wedge dz^{\beta}\wedge dz^{\gamma},\,\,
c^{(0,3)}=c_{\bar\alpha\bar\beta\bar\gamma}d\bar z^{\bar\alpha}\wedge d\bar z^{\bar\beta}\wedge d\bar z^{\bar\gamma},
&\nonumber\\
&c^{(2,1)}=3c_{\alpha\beta\bar\gamma}dz^\alpha\wedge dz^{\beta}\wedge d\bar z^{\bar\gamma},\,\,
c^{(1,2)}=3c_{\alpha\bar\beta\bar\gamma}d z^{\alpha}\wedge d\bar z^{\bar\beta}\wedge d\bar z^{\bar\gamma},
&\lb{decomp}
\end{eqnarray}
We may then rewrite the constraints (\ref{torco}) in terms of forms
\begin{equation}
c^{(2,1)}=-\partial\Omega,\,\,\, c^{(1,2)}=\bar\partial\Omega.\label{chat}
\end{equation}
The 3-forms $c^{(p,q)}$ are also convenient objects to deal with the constraints (\ref{diffi}). Indeed, a short calculation leads to
\begin{equation}
\iota_{I}dc-\frac{2}{3}d\iota_{I}c=i(2\partial c^{(3,0)}+\frac{4}{3}\partial c^{(2,1)}-\frac{2}{3}\bar\partial c^{(2,1)}
+\frac{2}{3}\partial c^{(1,2)}-\frac{4}{3}\bar\partial c^{(1,2)}-2\bar\partial c^{(0,3)}).
\end{equation}
An important feature of the expression in the right-hand side is that $\bar\partial c^{(3,0)}$
and $\partial c^{(0,3)}$ do not appear. The constraints (\ref{diffi}) may now be written as
\begin{eqnarray}&
\partial c^{(3,0)}=0,\quad\bar\partial c^{(0,3)}=0,&\nonumber\\ & \partial c^{(2,1)}=0\,\,\bar\partial c^{(1,2)}=0,\,\,
 \partial c^{(1,2)}-\bar\partial c^{(2,1)}=0.&\label{chot}
 \end{eqnarray}
The constraints on the second line of equations (\ref{chot}) are automatically solved by the expressions in (\ref{chat}).
So it remains simply to solve the constraints of the first line as follows
\begin{equation}
c^{(3,0)}=\partial B^{(2,0)},\quad c^{(0,3)}=\bar\partial B^{(0,2)}.\label{chit}
\end{equation}
Or, in components,
\begin{equation}
c_{\alpha\beta\gamma}=B_{[\alpha\beta,\gamma]}, \quad c_{\bar\alpha\bar\beta\bar\gamma}
=B_{[\bar\alpha\bar\beta,\bar\gamma]}.\label{chitc}
\end{equation}

To summarize, the geometry of the target manifold is determined by the hermitian metric $g_{\alpha\bar\beta}$ (or by the $(1,1)$ form
$\Omega$ in (\ref{metfo})) and by the antisymmetric tensors $B_{\alpha\beta}$ and $B_{\bar\alpha\bar\beta}$
(or by the forms $B^{(2,0)}$ and $B^{(0,2)}$). The torsion is then determined using (\ref{chat}) and (\ref{chit}).
Notice that a change in the 2-forms (i.e., a target space gauge transformation)
\begin{equation}
B^{(2,0)}\,\longrightarrow B^{(2,0)}+\partial V^{(1,0)},\quad
B^{(0,2)}\,\longrightarrow B^{(0,2)}+\bar\partial V^{(0,1)},\label{chau}
\end{equation}
does not affect the geometry.

\subsection{${\cal N}=2$ superspace}
We now consider things the other way around, and construct the most general action in ${\cal N}=2$ superspace
with coordinates $t,\theta,\bar\theta$. The supersymmetric covariant derivatives are
\begin{equation}
D=\frac{1}{\sqrt 2}\left(\frac{\partial}{\partial\theta}+i\bar\theta\partial_{t}\right),\,\,
\bar D=\frac{1}{\sqrt 2}\left(\frac{\partial}{\partial\bar\theta}+i\theta\partial_{t}\right),\,\, D^2=\bar D^2=0, \{D,\bar D\}=i\partial_{t}.
\end{equation}

We have to use chiral $Z^{\alpha}$, $\bar DZ^{\alpha}$ and antichiral $\bar Z^{\bar\alpha}$, $D\bar Z^{\bar\alpha}$,
superfields. The general form of the ${\cal N}=2$ supersymmetric action then reads \cite{Hu}
\begin{equation}
S=\frac{1}{2}\int dtd\theta d\bar\theta\,[-g_{\alpha\bar\beta}(Z,\bar Z)DZ^{\alpha}\bar D\bar Z^{\bar\beta}
+\frac{1}{12}B_{\alpha\beta}(Z,\bar Z)DZ^{\alpha}DZ^{\beta}
+\frac{1}{12}B_{\bar\alpha\bar\beta}(Z,\bar Z)\bar DZ^{\bar\alpha}\bar DZ^{\bar\beta}]\,.\label{n2act}
\end{equation}
It is naturally written in terms of the hermitian metric $g$ and 2-form $B$ which solve the geometrical
constraints in the previous subsection. Under a change of the $B$ tensors as in (\ref{chau}),
the integrand in the action $S$ is shifted by a total derivative
\begin{eqnarray}
& B_{\alpha\beta}\rightarrow B_{\alpha\beta}+V_{[\beta,\alpha]}, \,\,\,\,
B_{\bar\alpha\bar\beta}\rightarrow B_{\bar\alpha\bar\beta}+V_{[\bar\beta,\bar\alpha]}, &\nonumber\\&
S\rightarrow S+\int dtd\theta d\bar\theta \,[D(V_{\alpha}DZ^{\alpha})
+\bar D(V_{\bar\alpha}\bar D\bar Z^{\bar\alpha})]\,. &
\end{eqnarray}
We define the component fields by
\begin{equation}
z^{\alpha}=Z^{\alpha}\vert_{\theta=\bar\theta=0},\, \bar z^{\bar\alpha}=\bar Z^{\bar\alpha}\vert_{\theta=\bar\theta=0},\,
\lambda^\alpha=DZ^{\alpha}\vert_{\theta=\bar\theta=0},\, \bar \lambda^{\bar\alpha}
=\bar D\bar Z^{\bar\alpha}\vert_{\theta=\bar\theta=0}\,.
\end{equation}
When going to the component form of the action, one recovers an expression of the type (\ref{n1actc}),
with an hermitian metric $g_{\alpha\bar\beta}$, and the torsion is given by equations (\ref{torco})
and (\ref{chitc}). Thus, the geometrical constraints are automatically satisfied.

In Sect.7 we shall come back to this ${\cal N}=2$ superfield formalism to discuss general constraints imposed
by an extra ${\cal N}=2$ supersymmetry which builds up the manifest one to ${\cal N}=4$ supersymmetry.

\setcounter{equation}{0}

\section{${\cal N}{=}4\,, \; d{=}1$ harmonic superspace}
\subsection{Generalities}
We start this Section by recapitulating some basic facts about ${\cal N}{=}4$,
$d{=}1$ harmonic superspace (HSS),
following \cite{IL} and our papers \cite{gpr,DI2,DI3}.

The ordinary ${\cal N}{=}4, d{=}1$ superspace is defined as \be
(t,\theta_{i},\, \bar\theta^{i})\,,\quad \bar\theta^{i} =
\overline{(\theta_{i})}\,,\lb{N4} \ee where $t$ is the time
coordinate and the Grassmann-odd coordinates $\theta_{i},\,
\bar\theta^{i}$ form doublets of the  automorphism group
$SU(2)_{A}$\footnote{The second automorphism $SU(2)$ of the ${\cal
N}{=}4, d{=}1$ Poincar\'e supersymmetry is implicit in this notation;
it combines $\theta_i$ and $\bar\theta_i\,$ into a doublet.}. The
${\cal N}{=}4$ supertranslations act as \be \delta \theta_i =
\varepsilon_i\,, \quad \delta \bar\theta^i = \bar\varepsilon^i\,,\quad
\delta t = i \left(\bar\theta^i\varepsilon_i -
\bar\varepsilon^i\theta_i\right). \lb{N4susy} \ee The corresponding
covariant derivatives are
\begin{equation}
D^{i}=\frac{\partial}{\partial \theta_{i}}+i\bar\theta^{i}\partial_{t}\,,\;\;
\bar D_{i}=\frac{\partial}{\partial \bar\theta^{i}}+i\theta_{i}\partial_{t}\,, \quad
\{D^{i},\bar D_{j}\}=2i\delta^i_j\partial_{t}\,, \;\{D^{i}, D^{j}\}
= \{\bar D_{i},\bar D_{j}\} =0\,.
\end{equation}

${\cal N}{=}4, d{=}1$ HSS is an extension of \p{N4} by the harmonics
$u^\pm_i \in SU(2)_A/U(1)\,$. The basic relations the harmonics satisfy
are $u^{-}_i = \overline{(u^{+i})}\,$,
$u^{+i}u^-_{i}=1\,$. The latter constraint implies the completeness
relation
\be
u^+_iu^-_k - u^+_ku^-_i = \epsilon_{ik}\,. \lb{CompL}
\ee
The coordinates of ${\cal N}{=}4, d{=}1$ HSS
in the {\it analytic} basis are
\begin{equation}
\left( t_{A}=t-i(\theta^+\bar\theta^-+\theta^-\bar\theta^+)\,,\;
\theta^\pm=\theta^{i}u^\pm_{i}, \bar\theta^\pm=\bar\theta^{i}u^\pm_{i}, \;u^\pm_k \right).
\end{equation}
The analytic subspace of HSS is defined as the subset
\begin{equation}
(t_{A},\theta^+,\bar\theta^+, u^\pm_{i})\equiv (\zeta, u).
\end{equation}
It is closed under the ${\cal N}{=}4$ supersymmetry \p{N4susy}.

In the central basis $(t, \theta_i, \bar\theta^k, u^{\pm i})$ we define
the harmonic derivatives
and the harmonic projections of spinor derivatives as
\be
D^{\pm\pm} = \partial^{\pm\pm} = u^{\pm}_i\frac{\partial}{\partial u^{\mp}_i}\,,
\quad D^\pm=u^\pm_{i}D^{i},\quad \bar D^\pm=u^\pm_{i}\bar D^{i}\,,. \lb{Cderiv}
\ee
In the analytic basis, the same spinor and harmonic derivatives read
\begin{eqnarray}
&& D^+=\frac{\partial}{\partial\theta^{-}},\,\, \bar D^+=
-\frac{\partial}{\partial\bar\theta^-},\,\,
D^{-}=-\frac{\partial}{\partial \theta^{+}}+2i\bar\theta^{-}\partial_{t_{A}},\,\,
\bar D^{-}=\frac{\partial}{\partial \bar\theta^{+}}+2i\theta^{-}\partial_{t_{A}}\,,\nn
&& D^{++}=\partial^{++}-2i\theta^+\bar\theta^+\partial_{t_{A}}
+\theta^+\frac{\partial}{\partial\theta^-}
+ \bar\theta^+\frac{\partial}{\partial\bar\theta^-}\,, \nn
&& D^{--}=\partial^{--}-2i\theta^-\bar\theta^-\partial_{t_{A}}
+\theta^-\frac{\partial}{\partial\theta^+}
+  \bar\theta^-\frac{\partial}{\partial\bar\theta^+}\,.
\end{eqnarray}
The basic relations of the algebra of these derivatives are
\begin{eqnarray}
&&[D^{\pm\pm},D^\mp]=D^\pm,\,\, [D^{\pm\pm},\bar D^\mp]=\bar D^\pm,\,\,
\{D^+,\bar D^-\}=-\{D^-,\bar D^+\}=2i\partial_{t_{A}},  \nn
&& [D^{++},D^{--}]= D^{0}\,, \quad [D^0, D^{\pm\pm}] =
\pm 2 D^{\pm\pm}\,, \lb{DharmAl}\\
&& D^0 = u^+_{i}\frac{\partial}{\partial u^+_{i}}-u^-_{i}
\frac{\partial}{\partial u^-_{i}}+
\theta^+\frac{\partial}{\partial \theta^+}
+\bar\theta^+\frac{\partial}{\partial \bar\theta^+}
-\theta^-\frac{\partial}{\partial \theta^-}
-\bar\theta^-\frac{\partial}{\partial \bar\theta^-}\,.\label{Dalg}
\end{eqnarray}

The derivatives $D^+$, $\bar D^+$ are short in the analytic basis,
which implies the existence of analytic ${\cal N}{=}4$ superfields $\Phi^{(q)}(\zeta, u)$
\be
D^+\Phi^{(q)} = \bar D^+\Phi^{(q)}=0 \quad \Rightarrow \quad \Phi^{(q)}
= \Phi^{(q)}(\zeta, u)\,,\lb{AnalPhi}
\ee
$q$ being the external harmonic $U(1)$ charge, $D^0\Phi^{(q)} = q\,\Phi^{(q)}\,$. This Grassmann harmonic
analyticity is preserved
by the harmonic derivative $D^{++}$ which commutes with $D^+$ and $\bar D^+$.

Finally, the measures of integration over the full HSS and its analytic subspace
are given, respectively, by
\begin{eqnarray}
&& dudtd^4\theta=dudt_{A}(D^-\bar D^-)(D^+\bar D^+)=\mu_{A}^{(-2)}(D^+\bar D^+),\nn
&& \mu_{A}^{(-2)}=dud\zeta^{(-2)}=dudt_{A}d\theta^+d\bar\theta^+=dudt_{A}(D^-\bar D^-).
\end{eqnarray}

\subsection{Linear multiplet (4, 4, 0) in HSS}
Most of the ${\cal N}{=}4$, $d{=}1$ multiplets with four fermions,
namely, the multiplets ${\bf (1, 4, 3)}$, ${\bf (3, 4, 1)}$, ${\bf (4, 4, 0)}$ and ${\bf (0, 4, 4)}$,
have a concise off-shell description in terms of the harmonic analytic ${\cal N}{=}4$ superfields
\footnote{The multiplets ${\bf (2, 4, 2)}$ also admit a formulation
in ${\cal N}{=}4$, $d{=}1$ HSS, though in some indirect way \cite{DI3}.}. They are known to exist in linear
and nonlinear versions \cite{GR1,Pol,IL,IKL1,gpr,ksch,bks}.
The multiplets  ${\bf (4, 4, 0)}$ can be considered as a ``root''
${\cal N}{=}4$, $d{=}1$ multiplets: as shown in \cite{root} on the component level and
in \cite{gpr}, \cite{DI2}, \cite{DI3} in the superfield approach, the invariant actions of the other
multiplets can be obtained from the general ${\bf (4, 4, 0)}$ action by some well-defined procedure.

The standard {\it linear} multiplet ${\bf (4,4,0)}$ is described by a doublet analytic superfield
$q^{+a}(\zeta,u)$
of charge $1$
satisfying the non-dynamical harmonic constraint \footnote{For brevity,
in what follows we frequently
omit the index ''A'' of $t_A$.}
\begin{equation}
D^{++}q^{+a}=0 \quad \Rightarrow \quad q^{+a}(\zeta,u)=f^{ia}(t)u^+_{i}+\theta^+\chi^{a}(t)
+\bar\theta^+\bar\chi^{a}(t)+2i\theta^+\bar\theta^+\partial_t f^{ia}(t)u^-_{i}.\label{coq}
\end{equation}
The Grassmann analyticity conditions together with the harmonic constraints \p{coq}
imply that in the central basis
\bea
&& q^{+ a} = q^{ia}(t,\theta, \bar\theta)u^{+}_i\,, \;\;
D^{(i}q^{k)a} = \bar D^{(i}q^{k)a} = 0\,. \label{qconsC}
\eea

We may write a general off-shell action for the linear ${\bf (4,4,0)}$ multiplet as
\begin{equation}
S_{q}=\int dudtd^4\theta \,{\cal L}(q^{+a}, q^{-b}, u^\pm) , \,\,\lb{qact}
\ee
where
\be
q^{-a}\equiv D^{--}q^{+a} = q^{ia}u^{-}_i.
\end{equation}
The action \p{qact} can be rewritten in terms of the ordinary ${\cal N}=4$ superfield $q^{ia}$ as
\begin{equation}
S_{q} =\int dtd^4\theta\, L(q^{ia})\,, \quad L(q^{ia})=\int du\,{\cal L}(q^{+a}, q^{-b}, u^\pm)\,.
\end{equation}
The free action is given by
\begin{equation}
S_{q}^{\mbox{\scriptsize free}} = -\frac{1}{8}\,\int dtd^4\theta\, q^{ia}q_{ia}
= -\frac{1}{4}\,\int dudtd^4\theta\,q^{+a}q^-_{a}
=\frac{i}{2}\int dud\zeta^{(-2)}\, q^{+a}\partial_t q^+_{a}. \lb{Freeq}
\end{equation}

The action \p{qact} produces a sigma-model type action in components. The corresponding target metric is necessarily
conformally flat in the case of one multiplet (i.e. for 4-dimensional target space) and is given by the expression
\be
g_{ia\,kb} \sim \Delta_{[a b]}\,L(f) \,\epsilon_{ik}, \quad \Delta_{[a b]} = \frac{\partial^2}{\partial f^{ja} \,\partial f_{j}^{b}}\,,
\quad L(f) = L(q^{ia})|_{\theta = 0}\,, \lb{cflatn}
\ee
for the general $4n$-dimensional target\footnote{In a complex parametrization of the target space,
the same form of the metric was derived in \cite{Strom,Hu}.}.

One can also construct an invariant which
in components yields a Wess-Zumino type action,
with one time derivative on the bosonic fields plus Yukawa-type fermionic terms.
It is given by
the following integral over
the analytic subspace
\be
S_{WZ} = i\int du d\zeta^{(-2)}\, {\cal L}^{+2}(q^{+a}, u^\pm)\,. \lb{WZq}
\ee
The component Lagrangian contains the Lorentz-force type coupling $\sim A_{ia}(f)\dot{f}^{ia}$, where the abelian external gauge field
$A_{ia}$ satisfies the $\mathbb{R}^4$ self-duality condition in the 4-dimensional case and the appropriate generalization of this condition
in the $4n$-dimensional case \cite{IL}.

\setcounter{equation}{0}
\section{Nonlinear $q^+$ multiplet and its off-shell action}
\subsection{Most general nonlinear (4,4,0) multiplets}
The idea of nonlinear ${\cal N}{=}4, d=1$ $q^+$ multiplet is based on the analogy of the kinematical harmonic constraint \p{coq}
with the superfield equation
of motion of the free $q^+$ hypermultiplet in ${\cal N}{=}2, d=4$ HSS \cite{HSS,HSS1}. The most general
self-interaction of $n$ hypermultiplets $q^{+ A}, A =1, \ldots 2n$, at the level of the superfield equations of motion,
corresponds just to the replacement
\be
\mbox{\bf (a)}\; D^{++} q^{+ A} = 0\quad \Rightarrow \quad \mbox{\bf (b)} \; D^{++} q^{+ A} =
\Omega^{[AB]}\, \frac{\partial {\cal L}^{+ 4}}{\partial q^{+ B}}\,. \lb{hk}
\ee
Here, ${\cal L}^{+ 4} = {\cal L}^{+ 4}(q^+, u^\pm)$ is the harmonic superspace hyper-k\"ahler potential which encodes the full information about
the bosonic hyper-k\"ahler sigma model Lagrangian appearing as the purely bosonic part of the component hypermultiplet Lagrangian
\cite{hkHSS,geom1}\footnote{This harmonic hyper-K\"ahler  potential is a direct analog of the K\"ahler potential appearing as the most general
Lagrangian of ${\cal N}{=}1, d{=}4$ chiral superfields \cite{Zum}.}, and $\Omega^{[AB]}$ is a constant skew-symmetric $USp(2n)$ metric
($\Omega^{[AB]}\Omega_{[BC]} = \delta^A_C\,$).

In our $d=1$ case, the constraint \p{coq} is analogous to the free $d=4$ hypermultiplet equation of motion (\ref{hk}a), but its crucial
difference from the latter is that it just reduces the infinite field contents of the harmonic analytic ${\cal N}{=}4, d=1$
superfield $q^{+ a}$ to the irreducible component set ${\bf (4, 4, 0)}$, yet entailing no dynamical equations for
the remaining fields. Then an obvious nonlinear generalization of the harmonic constraint \p{coq} is
\be
D^{++}q^{+ a} = {\cal L}^{+3 a}(q^+, u^\pm)\,, \lb{coq-n}
\ee
where ${\cal L}^{+3 a}$ is an arbitrary charge 3 function of its arguments. This generalized constraint preserves the Grassmann
harmonic analyticity and supersymmetry. Any higher-order derivative $(D^{++})^p q^{+ a}$ is expressed through $q^{+ a}$ from \p{coq-n}.
Also, as in the linear case, the only other independent harmonic derivative of $q^{+ a}$ is the non-analytic one $D^{--}q^{+ a}$:
any higher-order derivative
$(D^{--})^p q^{+a}, \;p>1\,$, can be shown to satisfy, as a consequence of \p{coq-n}, a harmonic differential equation which
expresses it (at least perturbatively) in terms of $q^{+ a}$ and $D^{--}q^{+ a}$. In other words, as in the linear case, the only
independent harmonic projections of $q^{+ a}$ are this superfield itself and the non-analytic superfield $q^{- a} = D^{--}q^{+ a}$.

In what follows we assume that the index $a$ runs from 1 to $2n$, which corresponds to considering $n$ ${\bf (4,4,0)}$
multiplets. For the component fields in the $\theta$-expansion of $q^{+ a}$,
\be
q^{+ a}(\zeta, u) = f^{+ a}(t,u) + \theta^+ \,\chi^a(t, u) + \bar\theta^+\,\bar\chi^a (t, u) + \theta^+\bar\theta^+ \,A^{- a}\,,\lb{Expq+}
\ee
the superfield constraint \p{coq-n} implies the following harmonic constraints
\bea
&& \partial^{++} f^{+ a} = {\cal L}^{+3 a}(f^+, u^\pm) \,, \lb{Bas} \\
&& {\cal D}^{++} \chi^a = {\cal D}^{++} \bar\chi^a = 0\,, \lb{Spin} \\
&& {\cal D}^{++} A^{- a} = 2i \dot{f}^{+ a} +E^{+ a}_{\;cd}\bar\chi^c\chi^d\,, \lb{A-}
\eea
where
\bea
&& {\cal D}^{++} G^a(t, u) := \partial^{++}G^a - E^{+ 2 a}_{\;\;\;b} G^b\,, \quad {\cal D}^{++} F_a(t, u) :=
\partial^{++}F_a + E^{+ 2 b}_{\;\;\;a} F_b\,, \lb{dEf1} \\
&& E^{+ 2 a}_{\;\;\;b} (f^+,u) := \frac{\partial{\cal L}^{+3 a}}{\partial f^{+ b}}\,, \lb{dEf2}
\eea
for any contra- and covariant $2n$-vectors $G^a$ and $F_a\,$, and
\be
E^{+ a}_{\;bc} = E^{+ a}_{\;cb} := \partial_{+ b}E^{+2 a}_{\;\;\;c} = \partial_{+ b}\partial_{+ c}{\cal L}^{+3 a}\,. \lb{DefEabc}
\ee
The role of the constraints \p{Bas} - \p{A-} is the same as
that of the kinematical part of the $d=4$ constraints \p{hk}: they fix the harmonic dependence of all involved fields, expressing
them through the ordinary fields $f^{ia}(t) (i=1,2), \chi^a(t)$ and $\bar\chi^a(t)$ which appear as the integration constants in these
1-st order harmonic differential equations. It is seen that \p{Bas} - \p{A-} do not imply equations of motion for the physical
fields. An important corollary of \p{Bas} is
\be
{\cal D}^{++} \dot{f}{}^{+ a} = 0\,. \lb{Corol1}
\ee

\subsection{Geometric considerations}
As was already mentioned, the only independent harmonic projection of $q^{+ a}$, besides $q^{+ a}$ itself, is $q^{-a} := D^{--}q^{+a}$.
When expanded in $\theta, \bar\theta $, the non-analytic superfield $q^{-a}$ reads
\bea
q^{-a} = D^{--}q^{+ a} &=& \partial^{--}f^{+a} - 2i\theta^-\bar\theta^-\,\dot{f}{}^{+a} + \theta^-\,\chi^a + \bar\theta^-\,\bar\chi^a +
\theta^+\,\partial^{--}\chi^a + \bar\theta^+\,\partial^{--}\bar\chi^a \nn
&&+\; \theta^+\bar\theta^+\,\partial^{--}A^{-a} +(\theta^-\bar\theta^+ +\theta^+\bar\theta^-)\, A^{-a}
-2i\theta^-\bar\theta^-\theta^+\,\dot{\chi}{}^a \nn
&& -2i\theta^-\bar\theta^-\bar\theta^+\,\dot{\bar\chi}{}^a-2i\theta^-\bar\theta^-\theta^+\bar\theta^+\,\dot{A}{}^{-a}\,. \lb{q-dec}
\eea
We can choose as the second half of the target space bosonic coordinates the first component in \p{q-dec}, i.e. the quantity
\be
f^{-a} := \partial^{--} f^{+ a}\,. \lb{Deff-}
\ee
The set of the target space coordinates $\{f^{+ a}(t, u), f^{- a}(t, u), u^{\pm i}\}$ define a ``$\lambda$ basis''
in the harmonic extension $\{f^{ia}(t), u^{\pm i}\}$ of the target bosonic space $\{f^{ia}(t)\}$. The latter parametrization
is called the ``$\tau$ basis'' (see \cite{HSS,HSS1,dks} for the similar terminology in the case of the harmonic description of hyper-K\"ahler
geometry). The coordinates in both bases are related through the ``bridges'':
\be
f^{+ a} = f^{ia}u^+_i + v^{+ a}(f^{ia}, u^\pm)\,, \quad f^{- a} = f^{ia}u^-_i + v^{- a}(f^{ia}, u^\pm)\,, \; v^{- a} = \partial^{--}v^{+ a}\,,
\lb{Bridges}
\ee
the basic relation \p{Bas} being interpreted as the equation specifying the bridge $v^{+ a}$ in terms of the $\tau$ basis coordinates
\be
\partial^{++}v^{ +a} = {\cal L}^{+3 a}(f^{ia}u^+_i + v^{+ a}, u^\pm)\,.
\ee
It will be useful to give the harmonic derivatives in the considered $\lambda$ basis:
\bea
\partial^{++} &=& \hat{\partial}^{++} +\partial^{++}f^{+a}\frac{\partial}{\partial f^{+ a}}
+ \partial^{++}f^{-a}\frac{\partial}{\partial f^{- a}} \nn
&=&
\hat{\partial}^{++} + {\cal L}^{+3 a}\frac{\partial}{\partial f^{+ a}} + \left(f^{+a} + \hat{\partial}^{--} {\cal L}^{+3 a}
+f^{-b}E^{+2 a}_{\;\;\;b} \right)\frac{\partial}{\partial f^{- a}}\,, \lb{++lAmbda} \\
\partial^{--} &=& \hat{\partial}^{--} + f^{-a}\frac{\partial}{\partial f^{+ a}}
+ \partial^{--}f^{-a}\frac{\partial}{\partial f^{- a}}\,, \lb{--lAmbda}
\eea
where $\hat{\partial}^{\pm\pm}$ denote those parts of the harmonic derivatives which act only on the explicit harmonics (and
so do not act on $f^{\pm a}$) and we made use of the relations $[\partial^{++}, \partial^{--}] = \partial^0$ and \p{Bas}, \p{Deff-}.
Note that \p{Bas} implies the harmonic differential equation for $\partial^{--}f^{-a}$:
\be
{\cal D}^{++}(\partial^{--}f^{-a}) = ( \hat{\partial}^{--})^2{\cal L}^{+3 a} + 2f^{-b}\hat{\partial}^{--}E^{+2 a}_{\;\;\;b}
+ f^{-b}f^{-c} E^{+ a}_{\;bc}\,. \lb{--f-}
\ee
It follows from eq. \p{--f-} that $\partial^{--}f^{-a} = (\partial^{--})^2f^{+a}$ (and all higher-order $\partial^{--}$
derivatives of $f^{+ a}$) are indeed expressed through $f^{+ a}$ and $f^{-a}$ as the only independent coordinates of
the target manifold in the $\lambda$-basis.

The relations \p{Bas} - \p{A-} and \p{Corol1} are covariant under the analytic target space diffeomorphism transformations
\bea
&& \delta f^{+ a} = \lambda^{+a}(f^+, u)\,, \;\delta {\cal L}^{+3 a} = {\cal L}^{+3 b}\partial_{+ b}\lambda^{+ a}
+ \hat{\partial}{}^{++}\lambda^{+ a}\,, \lb{Diff1} \\
&&\delta f^{-a} = \hat{\partial}{}^{--}\lambda^{+ a} + f^{-b}\partial_{+b}\lambda^{+a}\,, \lb{f-Tran}  \\
&& \delta E^{+2 a}_{\;\;\;b} = -\partial_{+ b}\lambda^{+ c} E^{+2 a}_{\;\;\;c} + E^{+2 d}_{\;\;\;b}\partial_{+ d}\lambda^{+ a}
+ \partial^{++}(\partial_{+ b}\lambda^{+ a})\,, \lb{Diff2} \\
&& \delta \chi^a = \partial_{+ b}\lambda^{+a}\chi^b\,, \; \delta \bar\chi^a = \partial_{+ b}\lambda^{+a}\bar\chi^b\,, \; \nn
&& \delta A^{-a} = \partial_{+ b}\lambda^{+a}A^{-b} + \partial_{+ c}\partial_{+ d}\lambda^{+ a}\,\bar\chi{}^c\chi^{d}\,. \;\lb{Atransf}
\eea
Note that the transformation laws \p{f-Tran} and \p{Diff2} follow from \p{Diff1} and the definitions \p{dEf2}, \p{Deff-}.
As we see, the target analyticity-preserving diffeomorphisms induce the analytic tangent $GL(2n)$ transformations\footnote{Strictly speaking,
the tangent space gauge group is $GL(n, \mathbb{H}) \supset USp(2n)\,;$ for simplicity, we denote it $GL(2n)$,
hoping that this will not result in a misunderstanding.} of the external tensor indices with the gauge parameter $\partial_a\lambda^b(f^+,u)$,
i.e. in the present case
we are dealing with a gauge-fixed version of the $\lambda$ basis target geometry, like in the HSS formulations of
the hyper-K\"ahler \cite{geom1}, quaternion-K\"ahler \cite{qkHSS} and the ${\cal N}{=}4$ heterotic sigma models
geometries \cite{dks}.

For what follows it is useful to introduce a non-analytic
harmonic vielbein $E^{-2 a}_{\;\;\;b}(f^+, f^-, u)$ related to $E^{+2 a}_{\;\;\;b}(f^+, u)$ by the harmonic zero-curvature
equation
\be
\partial^{--}E^{+2 a}_{\;\;\;b} = {\cal D}^{++}E^{-2 a}_{\;\;\;b}\,. \lb{ZC2}
\ee
The covariance of this equation requires that $E^{-2 a}_{\;\;\;b}$ transforms in the following way
\be
\delta E^{-2 a}_{\;\;\;b} = -\partial_{+ b}\lambda^{+ c} E^{-2 a}_{\;\;\;c} + E^{-2 d}_{\;\;\;b}\partial_{+ d}\lambda^{+ a}
+ \partial^{--}(\partial_{+ b}\lambda^{+ a})\,. \lb{Diff3}
\ee
This quantity is just the harmonic connection covariantizing the derivative $\partial^{--}$:
\be
{\cal D}^{--} G^a(t, u) := \partial^{--}G^a - E^{- 2 a}_{\;\;\;b} G^b\,, \quad {\cal D}^{--} F_a(t, u) :=
\partial^{--}F_a + E^{- 2 b}_{\;\;\;a} F_b\,. \lb{dEf22}
\ee
In virtue of the condition \p{ZC2} these covariantized harmonic derivatives satisfy the same algebra as the flat ones
\be
[{\cal D}^{++}, {\cal D}^{--}] = \partial{}^0\,. \lb{CovDD}
\ee

The fact that there exist the analytic frame $GL(2n)$ transformations implies the necessity of introducing, apart from the bridges
between $\lambda$ and $\tau$ bases, also some $GL(2n)$ valued
``bridges'' $M_a^{\underline{b}}$ which relate the analytic $\lambda$ frame to the $\tau$ frame in which gauge transformations do not depend
on harmonics. These frame bridges are defined by the relations
\bea
&&\partial^{\pm\pm}M_a^{\underline{b}} + E^{\pm 2 c}_{\;\;\;a}M_c^{\underline{b}} = 0\,, \quad
\partial^{\pm\pm}(M^{-1})_{\underline{b}}^a
- E^{\pm 2 a}_{\;\;\;c}(M^{-1})^c_{\underline{b}} = 0\,, \lb{BridEq} \\
&& M_a^{\underline{b}}(M^{-1})^a_{\underline{c}}
= \delta^{\underline{b}}_{\underline{c}}\,, \quad M_a^{\underline{b}}(M^{-1})^b_{\underline{b}} = \delta^{b}_{a}\,, \nn
&& \delta M_a^{\underline{b}} = -\partial_{+ a}\lambda^{+c}M_c^{\underline{b}}\,, \quad
\delta(M^{-1})^a_{\underline{c}} = \partial_{+b}\lambda^{+ a}(M^{-1})^b_{\underline{c}}\,. \lb{FrBrid}
\eea
The underlined indices are inert under the $\lambda$ frame group, on them some harmonic-independent $\tau$ gauge group
is appropriately implemented. This $\tau$ gauge group can be interpreted as a redundancy in solving the harmonic equations \p{BridEq}
which express (non-locally in harmonics) the bridges in terms of $E^{\pm c}_a$:
\bea
M^{\und{b}}_a =  \tilde{M}^{\und{d}}_a\,L^{\und{b}}_{\und{d}}\,, \;\, (M^{-1})_{\underline{b}}^a =
(L^{-1})^{\und{d}}_{\underline{b}}\,(\tilde{M}^{-1})_{\underline{d}}^a \,,  \;\,
\partial^{\pm \pm}L^{\und{b}}_{\und{d}} = 0\,, \;\, \tilde{M}^{\und{d}}_a =
\delta^{\und{d}}_a - (\partial^{++})^{-1}\,E^{+ 2 \und{d}}_a +\ldots, \lb{Redund}
\eea
where $(\partial^{++})^{-1}$ is the appropriate harmonic Green function \cite{HSS1}. The harmonic-independent $GL(2n)$ gauge group matrices
$L^{\und{b}}_{\und{d}}$ can be used to fix one or another convenient $\tau$ gauge.

Using the frame bridges, one can relate various $\lambda$ and $\tau$ frame tensors. In particular,
the objects in the $\lambda$ frame subjected to the covariant conditions of harmonic independence, e.g.
fermions $\chi^a, \bar\chi^a$ satisfying \p{Spin}, can be transformed into objects which are manifestly
harmonic-independent, and {\it vice versa}:
\be
\chi^{\underline{a}} = M_b^{\underline{a}}\chi^b\,, \; \bar\chi^{\underline{a}} = M_b^{\underline{a}}\bar\chi^b\,,
\quad \mbox{eq. \p{Spin}} \;\Leftrightarrow \; \partial^{++}\chi^{\underline{a}} = \partial^{++}\bar\chi^{\underline{a}} = 0\,.\lb{Hindchi}
\ee
Also, exploiting bridges allows one to show the validity of a covariantized version of the well-known lemma \cite{HSS,HSS1}
\be
{\cal D}^{++} F^{ q\, a_1 \cdots a_p}_{\;\;\;b_1 \cdots b_k} = 0 \; \Rightarrow \;
F^{ q\, a_1 \cdots a_p}_{\;\;\;b_1 \cdots b_k} = 0 \quad \mbox{for $q < 0$} \lb{lemma}
\ee
and of the similar proposition with ${\cal D}^{--}$ (for $q> 0$).

The last geometric objects of the $\lambda$ geometry which we shall need are the covariant derivatives with respect to
the analytic basis coordinates
\be
{\cal D}_{+a}G^b = \nabla_{+ a}G^b - E^{- b}_{\;ac}\,G^c\,,\; {\cal D}_{+a}F_b = \nabla_{+ a}F_b
+ E^{- c}_{\;ab}\,F_c\,, \quad {\cal D}_{-a} = \partial_{-a}\,, \lb{DerCo}
\ee
where
\be
\nabla_{+ a} = \partial_{+ a} + E^{-2 b}_{\;\;\;a}\,\partial_{-b}\,, \quad {\cal D}^{++}E^{-a}_{\;cd} = E^{+ a}_{\;cd}\,. \lb{DerCo2}
\ee
The derivative $\partial_{-a}$ does not need to be covariantized due to the analyticity of $\lambda^{+ a}\,,
\;\partial_{-b}\lambda^{+ a} =0\,$. {}From the definition of $E^{-a}_{\,cd}$ and the proposition \p{lemma} it follows
that
\be
E^{-a}_{\,cd} = E^{-a}_{\,dc}\,.
\ee
Also, it is easy to check that
\be
{\cal D}^{++}\,\partial_{-b} E^{-2 a}_{\;\;\;d} = E^{+ a}_{\;bd}\,,
\ee
which, together with the definition \p{DerCo2}, by the lemma \p{lemma} implies
\be
E^{-a}_{\,cd} = \partial_{-c} E^{-2 a}_{\;\;\;d} = \partial_{-d} E^{-2 a}_{\;\;\;c}\,. \lb{Rel2}
\ee

The covariant derivatives defined above have the following commutation relations among themselves and with the covariantized harmonic derivatives
\bea
&& [{\cal D}_{+a},{\cal D}_{+b}] =[{\cal D}_{-a},{\cal D}_{-b}] = 0\,, \quad [{\cal D}_{+a},{\cal D}_{-b}] = {\cal R}^{+-}_{(ab)}\,,
\lb{CommA} \\
&& [{\cal D}^{++},{\cal D}_{-b}] = [{\cal D}^{--},{\cal D}_{+b}] = 0\,, \quad [{\cal D}^{++},{\cal D}_{+b}] = - {\cal D}_{-b}\,, \;
[{\cal D}^{--},{\cal D}_{-b}] = - {\cal D}_{+b}\,. \lb{CommB}
\eea
Here
\be
({\cal R}^{+-}_{(ab)})^c_d = \partial_{-a}\partial_{-b}E^{-2\,c}_{\;\;\;d} = \partial_{-a}E^{-\,c}_{\;\;bd}\,. \lb{Curvat}
\ee
This curvature is totally symmetric in the lower case indices as a consequence of \p{Rel2}.

Using the definitions and relations given above, as well as the general proposition \p{lemma}, one can derive a set of useful
identities to be exploited in what follows
\bea
&& {\cal D}^{--} E^{+a}_{\;bc} = E^{- a}_{\;bc} + {\cal D}^{++}\nabla_{+ c}E^{-2 a}_{\;\;\;b}\,, \lb{Id1} \\
&& {\cal D}^{--} E^{-a}_{\;bc} = \nabla_{+c}E^{-2a}_{\;\;\;b}\,, \lb{Id2} \\
&& {\cal D}^{--}\dot{f}{}^{+a} = \dot{f}{}^{-a} -E^{-2 a}_{\;\;\;b}\dot{f}{}^{+b}\,, \lb{Id3} \\
&& {\cal D}^{++}\left(\dot{f}{}^{-a}-E^{-2 a}_{\;\;\;b}\dot{f}{}^{+b}\right) = \dot{f}{}^{+a}\,,\lb{Id4} \\
&& {\cal D}^{--}\left(\dot{f}{}^{-a}-E^{-2 a}_{\;\;\;b}\dot{f}{}^{+b}\right) = ({\cal D}^{--})^2 \dot{f}{}^{+a} = 0\,.\lb{Id5}
\eea

These identities imply some corollaries. E.g., it follows from
\p{Id2} that
\be
\nabla_{+[c}E^{-2 a}_{\;\;\;b]} = 0 \quad\Rightarrow \quad [\nabla_{+c}, \nabla_{+b}] = 0\,. \lb{Rel3}
\ee
Also, comparing the second of eqs. \p{DerCo2} and eq. \p{Id4} with
the constraint \p{A-} on $A^{-a}$ and taking into account the
constraints \p{Spin}, we observe that $A^{-a}$ can be represented as
\be
A^{-a} = \hat{A}{}^{-a} + E^{-a}_{\;bc}\,\bar\chi^b\chi^c\,,
\quad \hat{A}{}^{-a} = 2i\left(\dot{f}{}^{-a}-E^{-2a}_{\;\;\;b}\dot{f}{}^{+b}  \right), \;\;
{\cal D}^{--}\hat{A}{}^{-a}=0\,. \lb{A-expr}
\ee
Another useful identity
is
\be
\nabla_{+[c}E^{-a}_{\;d]b} + E^{-f}_{\;b[c}E^{-a}_{\;d]f} =
0\,. \lb{Rel4}
\ee
One more corollary is the relation
\be
E^{-2\,a}_{\;\;\;b} = \frac{1}{2}\,\partial_{-b}(\partial^{--}f^{-a})\,.
\ee

Finally, note that we could restrict the tangent frame gauge group from $GL(2n)$ to its subgroup $USp(2n)\subset GL(2n)$.
This restriction amounts to the requirement of preserving a constant skew-symmetric tensor $\Omega_{[a\,b]}$:
\be
\partial_{+ a}\lambda^{+c}\Omega_{[cb]} + \partial_{+b}\lambda^{+c}\Omega_{[ac]} = 0 \;\Rightarrow \; \partial_{+[a}\lambda^+_{c]} = 0\,, \;\;
\lambda^+_a := \Omega_{[ab]}\lambda^{+ b}\,. \lb{gltosp}
\ee
This implies
\be
\lambda^+_b = \partial_{+b}\lambda^{++}(f^+,u)\,, \;\; \partial_{+ a}\lambda^{+b}= \Omega^{[bc]}\partial_{+ a}\partial_{+c}\lambda^{++}
\equiv \Omega^{[bc]}\lambda_{(ac)}\,, \lb{spnparam}
\ee
where $\lambda^{++}$ is an arbitrary charge 2 analytic harmonic function. After restricting to $USp(2n)$ as the frame gauge group, it
is consistent to impose the $USp(2n)$ invariant constraint
\be
\partial_{+[a}{\cal L}^{+3}_{b]} = 0\,, \quad {\cal L}^{+3}_{b} := \Omega_{[bc]}{\cal L}^{+3c}\,, \lb{Constr}
\ee
which is solved by
\be
{\cal L}^{+3}_{a} = \partial_{+a}{\cal L}^{+4}(f^+, u)\,. \lb{HK}
\ee
The gauge transformation law \p{Diff1} restricted to the $USp(2n)$ gauge group with the infinitesimal parameters \p{spnparam}
implies for the potential  ${\cal L}^{+4}$ the simple transformation law
\be
\delta {\cal L}^{+4} = \hat{\partial}{}^{++}\lambda^{++}\,. \lb{L4tran}
\ee

It should be pointed out that there is no intrinsic reason to impose the constraint \p{HK}. It {\it can be} imposed if
the analytic tangent frame group is restricted from $GL(2n)$ to $USp(2n)$. Nevertheless, as will be clear soon (Sect. 6),
this extra constraint has a simple geometric interpretation.

\subsection{The invariant actions}
As the independent superfield projections of $q^{+a}$ in the nonlinear case are $q^{+a}$ and $q^{-a} = D^{--}q^{+a}$, like
in the case of linear harmonic constraints, the most general sigma model action  and the WZ type action look
like their linear case prototypes \p{qact} and \p{WZq}
\bea
\tilde{S}_q = \int dt du d^4\theta\, {\cal L}(q^{+a}, q^{-b}, u^{\pm})\,, \lb{nonlqAct}
\eea
\bea
\tilde{S}_{WZ} = i\int du\zeta^{(-2)}\,{\cal L}^{+ 2}(q^{+a}, u^{\pm})\,. \lb{WZ2}
\eea
However, the component structure of these superfield actions radically differs from that of their linear analogs.

Let us first work out the component form of \p{nonlqAct}.
Substituting there the $\theta$ expansions \p{Expq+} and \p{q-dec}
and performing the integration over $\theta$s, we can rewrite
\p{nonlqAct} as
\be \tilde{S}_q = \int dt du \,{\cal L}^{comp}(f^{\pm}(t,u), \chi(t,u), \bar\chi(t,u), u^\pm).
\lb{nonlqAct2}
\ee
After rather cumbersome calculations making use
of the relations derived in the previous Subsection, in particular,
the identities and relations \p{Id1} - \p{A-expr}, we find the
following relatively simple expression for $\tilde{\cal L}^{comp}$:
\bea
{\cal L}^{comp} &=& {\cal L}^{(1)}+ {\cal
L}^{(2)}+ {\cal L}^{(3)}\,, \lb{lagr} \\
{\cal L}^{(1)} &=&
4i{\cal F}_{[a b]}\left\{2i \dot{f}{}^{+[a}\left(\dot{f}{}^{-b]} -
E^{-2\,b]}_{\;\;\;c}\dot{f}{}^{+c}\right)
+ \frac{1}{2}\left(\nabla\bar\chi^{[a}\chi^{b]} - \bar\chi^{[a}\nabla\chi^{b]}\right)\right\} \lb{Lagr1} \\
{\cal L}^{(2)} &=&  4i\left\{{\cal D}_{+b}{\cal F}_{[ac]}\,\dot{f}{}^{+a} +
{\partial}_{-b}{\cal F}_{[ac]}\,\left(\dot{f}{}^{-a} - E^{-2\,a}_{\;\;\;d}\dot{f}{}^{+d}\right)\right\}  \bar\chi^{(b}\chi^{c)}\lb{Lagr2} \\
{\cal L}^{(3)} &=& \left({\cal D}_{+a}{\cal D}_{+b}{\partial}_{-c}{\partial}_{-d}{\cal L}(f,u)\right)
\bar\chi^{(a}\chi^{b)}\bar\chi^{(c}\chi^{d)}\nn
 &=& -\left({\cal D}_{+[b}{\partial}_{-d]}{\cal F}_{[a c]}\right)
\bar\chi^{a}\bar\chi^{c}\chi^{b}\chi^{d}\, = -\left({\partial}_{-[d}{\cal D}_{+b]}{\cal F}_{[a c]}\right)
\bar\chi^{a}\bar\chi^{c}\chi^{b}\chi^{d}\,. \lb{Lagr3}
\eea
Here
\bea
&&{\cal F}_{[ac]} = {\cal D}_{+[a}{\partial}_{-c]}{\cal L} = \nabla_{+[a}{\partial}_{-c]}\,{\cal L} = \partial_{-[c}\nabla_{+a]}\,{\cal L} \,,
\label{defF} \\
&& \nabla\chi^a = \dot{\chi}{}^a - \dot{f}{}^{+ d}E^{-\,a}_{\;d c}\chi^c\,, \quad
\nabla\bar\chi^a = \dot{\bar\chi}{}^a - \dot{f}{}^{+ d}E^{-\,a}_{\;d c}\bar\chi^c\,,\lb{defUr}
\eea
and
\be
{\cal L}(f, u) = {\cal L}(q^+, q^-,u^\pm)\vert_{\theta = \bar\theta =0}\,. \lb{DefL}
\ee
Note that the quartic term \p{Lagr3} is not changed under the permutation of the pairs of the lower case indices
$(d, b) \,\leftrightarrow \, (a,c)$ as a consequence of the fact that the curvature \p{Curvat} is totally symmetric in its lower case
indices. The symplectic ``metric'' ${\cal F}_{[a\,c]}$, as a consequence of its definition and some identities derived in Sect. 3.2,  satisfies
simple cyclic relations
\bea
\partial_{-a}{\cal F}_{[bc]} + \mbox{cycle} = 0\,, \quad
{\cal D}_{+ a}{\cal F}_{[bc]} + \mbox{cycle} = {\nabla}_{+ a}{\cal F}_{[bc]} + \mbox{cycle} = 0\,. \lb{Cycle2}
\eea

While deriving \p{Lagr1} - \p{Lagr3}, we started from the superfield Lagrangian, so the component action \p{nonlqAct} is ${\cal N}=4$
superinvariant by construction. A good check of the correctness of the component action is to explicitly show that it is invariant
under the off-shell ${\cal N}=4$ supersymmetry transformations of the component fields:
\bea
&&\delta f^{+ a} = \varepsilon^+\chi^a +\bar\varepsilon{}^+ \bar\chi{}^a\,, \quad \delta f^{-a} =
\varepsilon^-\chi^a +\bar\varepsilon{\,}^- \bar\chi{}^a
+ \varepsilon^+ E^{-2 \,a}_{\;\;\;b}\chi^b + \bar\varepsilon{}^+ E^{-2 \,a}_{\;\;\;b}\bar\chi{}^b\,, \nn
&&\delta \chi^a = \bar\varepsilon{}^+\left[2i\left(\dot{f}{}^{-a} - E^{-2\,a}_{\;\;\;b}\dot{f}{}^{+b}\right)
+ E^{-a}_{\;bc}\bar\chi{}^b\chi^c\right]
-2i\bar\varepsilon{\,}^-\dot{f}{}^{+a}\,, \nn
&&\delta \bar\chi^a = -\varepsilon^+\left[2i\left(\dot{f}{}^{-a} - E^{-2\,a}_{\;\;\;b}\dot{f}{}^{+b}\right)
+ E^{-a}_{\;bc}\bar\chi{}^b\chi^c\right]
+2i\varepsilon^-\dot{f}{}^{+a}\,, \lb{compN4}
\eea
where $\varepsilon^{\pm} = -\varepsilon^iu^{\pm}_i\,, \bar\varepsilon{}^{\pm} =
-\bar\varepsilon^iu^{\pm}_i$ and we made use of the expression \p{A-expr} for $A^{-a}$, as well as the relations
\be
{\cal D}^{--}\chi^a = \partial^{--}\chi^a - E^{-2\,a}_{\;\;\;b}\chi^b = 0\,, \quad
{\cal D}^{--}\bar\chi{}^a = \partial^{--}\bar\chi{}^a - E^{-2\,a}_{\;\;\;b}\bar\chi{}^b = 0\,,
\ee
which follow from \p{Spin}. Note that the ${\cal N}=4$ transformation of the composite component field $A^{-a}\,$
in the $\theta$-expansion \p{Expq+},
\be
\delta A^{-a} = 2i\varepsilon^{-}\dot{\chi}{}^a + 2i \bar\varepsilon^{-}\dot{\bar\chi}{}^a\,,
\ee
can now be reproduced as a result of varying \p{A-expr} with respect to the ${\cal N}=4$ supersymmetry transformations \p{compN4}.

After rather long computation we have checked that the total Lagrangian \p{lagr} is invariant under \p{compN4} modulo a total
time derivative, i.e. the action \p{nonlqAct} is indeed ${\cal N}=4$ supersymmetric. In this check, we essentially used the cyclic relations
\p{Cycle2}, as well as the relations \p{Rel2}, \p{Rel3} and \p{Rel4}.

Thus we constructed the most general ${\cal N}=4$ supersymmetric sigma-model action of the generic nonlinear ${\cal N}=4, d=1$ multiplet
${\bf (4,4,0)}$.
The relevant target geometry is fully specified by two prepotentials: the analytic potential ${\cal L}^{+3a}(f^+, u)$ and
general scalar non-analytic potential ${\cal L}(f^+, f^-, u)$. We will see in the next Section that this geometry is the so-called
weak hyper-K\"ahler geometry with torsion (weak HKT) \cite{hkt}. The hyper-K\"ahler (HK) geometry, as well as the ``strong'' HKT geometry,
are particular cases of this more general geometry. They correspond to some particular choices of the prepotentials just mentioned.

To complete the model, we now present the component form of the analytic superfield WZ term \p{WZ2}. It yields the most general
${\cal N}=4$ supersymmetric
coupling of the nonlinear ${\bf (4,4,0)}$ supermultiplet to the background abelian gauge field. Its precise form is found by performing in \p{WZ2}
the Berezin integration, and it is as follows
\bea
&& S_{WZ} = -\int dt du\, {\cal L}_{WZ}(t,u)\,,\lb{WZcomp1} \\
&& {\cal L}_{WZ} = 2 \left(\partial_{+a}{\cal L}^{+2}\right)\left(\dot{f}{}^{-a} - E^{-2 a}_{\;\;\;b}\dot{f}{}^{+b}\right) +
i\left({\cal D}_{+a}\partial_{+b}{\cal L}^{+2}\right)\chi^{(a}\bar\chi{}^{b)}\,. \lb{WZcomp}
\eea
Here ${\cal L}^{+2}(f^+, u) ={\cal L}^{+2}(q^+, u)\vert_{\theta^+ = \bar\theta^+ = 0}$.

The superfield WZ term \p{WZ2} is invariant under the following abelian  gauge target space transformation
\be
{\cal L}^{+2}{\,}'(q^+, u^\pm) = {\cal L}^{+2}(q^+, u) + D^{++}\sigma(q^+, u)\,.\lb{GaugeI}
\ee
The component realization of this superfield invariance is the invariance of \p{WZcomp} under the gauge transformation
\be
{\cal L}^{+2}{\,}'(f^+, u) = {\cal L}^{+2}(f^+, u) + \partial^{++}\sigma(f^+, u)\,, \quad
\sigma(f^+, u) =\sigma(q^+, u)\vert_{\theta^+ = \bar\theta^+ = 0}\,. \lb{GaugeCom}
\ee
Indeed, using the relation \p{Id4} and \p{Spin}, it is easy to check that ${\cal L}_{WZ}$ in \p{WZcomp} is shifted
by a total derivative under \p{GaugeCom},
\be
{\cal L}{}'_{WZ} = {\cal L}_{WZ}
+ \partial^{++}\left[2\partial_{+a}\sigma \left(\dot{f}{}^{-a} - E^{-2 a}_{\;\;\;b}\dot{f}{}^{+b}\right) +
i\left({\cal D}_{+a}\partial_{+b}\sigma\right)\,\chi^{(a}\bar\chi{}^{b)}\right] - 2\dot\sigma\,,
\ee
so the WZ action \p{WZcomp} is invariant. The Lagrangian \p{WZcomp} also transforms as a scalar under the target space
diffeomorphisms \p{Diff1} - \p{Atransf}, \p{Diff3}, assuming that ${\cal L}^{+2}(f^+, u)$ is a scalar, $\delta {\cal L}^{+2}(f^+, u) \simeq
{\cal L}^{+2}{}'(f^+{}', u) - {\cal L}^{+2}(f^+, u) = 0\,$.

The harmonic-independent form of the component Lagrangians corresponding to \p{lagr} and \p{WZcomp} will be presented in Sect. 5.7.

\setcounter{equation}{0}

\section{Geometry in the $\tau$ world}
\subsection{Vielbeins and complex structures}
Let us denote the harmonic-independent target coordinates in the
$\tau$ frame and $\tau$ basis by $f^{\,i b}, i = 1,2; b= 1,\ldots, 2n\,$. They are assumed to satisfy some reality condition, so we
deal with $4n$ real coordinates like in the hyper-K\"ahler sigma models. We define the $\tau$-frame vielbein
and its inverse by
\bea
&& E^{+ \underline{a}}_{\,i b} = \partial_{i b}f^{+ a}\, M^{\underline{a}}_a\,, \quad
E^{- \underline{a}}_{\,i b} = \left(\partial_{i b}f^{- a}
- E^{-2a}_{\;\;c}\,\partial_{i b}f^{+ c}\right)M^{\underline{a}}_a\,,\, \lb{Viel} \\
&& E_{+ \underline{a}}^{\,i b} = \nabla_{+ a} f^{i b}\,(M^{-1})^a_{\underline{a}}\,, \quad E_{- \underline{a}}^{\,ib}
= \partial_{- a}f^{i b}\,(M^{-1})^a_{\underline{a}}\,, \lb{VielInv} \\
&& \partial^{++}E^{+ \underline{a}}_{\,i b} = 0\,, \quad \partial^{++}E^{- \underline{a}}_{\,i b} = E^{+ \underline{a}}_{\,i b}\,\;
\Rightarrow \;E^{\pm \underline{a}}_{\,i b} = - e^{\,\underline{k}\underline{a}}_{\,i b}\,u^{\pm}_{\underline{k}}\,, \nn
&&\partial^{++}E_{- \underline{a}}^{\,ib } = 0\,,\quad \partial^{++}E_{+\underline{a}}^{\,i b} = -E_{- \underline{a}}^{\,i b}\,, \;
\Rightarrow \;E_{+\underline{a}}^{\,i b} = e_{\,\underline{k}\underline{a}}^{\,i b}u^{-\underline{k}}\,, \;
E_{-\underline{a}}^{\, i b} = -e_{\,\underline{k}\underline{a}}^{\,i b}u^{+\underline{k}}\,. \lb{Viel3}
\eea
The frame bridges $M^{\underline{a}}_a$ and $(M^{-1})^a_{\underline{a}}$ are defined in \p{FrBrid}. In \p{Viel3}, we underlined
the $\tau$ frame tangent space $SU(2)$ indices ($\und{i}, \und{k} = 1,2$) in order to distinguish them from the indices $i,k$
of the $\tau$ basis coordinates $f^{ia}$. The harmonic-independent $\tau$-frame vielbeins, as a consequence of their definition,
satisfy the following orthogonality conditions:
\be
e_{\,\underline{i}\underline{a}}^{\, j c}e^{\,\underline{k}\underline{b}}_{\, j c}
= \delta^{\underline{k}}_{\underline{i}}\,\delta^{\,\underline{b}}_{\,\underline{a}}\,, \quad
e_{\,\underline{k}\underline{a}}^{\,i b}e^{\,\underline{k}\underline{a}}_{\,j c} = \delta^i_j \,\delta^b_c \equiv \delta{}^{i b}_{j c}\,.
\ee

It is easy to define the appropriate triplet of complex structures. We combine the harmonic-independent fermionic fields $\chi^{\und{a}}$ and
$\bar{\chi}^{\und{a}}$ defined in \p{Hindchi}, as well the parameters
of the ${\cal N}=4$ supersymmetry transformations, into doublets of some extra $SU(2)\,$, i.e.
$(\chi^{\und{a}}, \bar{\chi}^{\und{a}}) \equiv \chi^{\und{a}\alpha}\,, \;(\varepsilon_i,
\bar\varepsilon_i) \equiv \varepsilon_{i\alpha}\,, \;\alpha =1,2\,$. Then we identify this $SU(2)$ with the one acting
on the indices of harmonic and tangent space $SU(2)$ indices $\und{i}, \und{k}$, and perform an equivalence transformation to the fermions
with the world indices $\chi^{ia}$ as
\be
\chi^{ia} = e^{\,ia}_{\und{k}\und{a}}\chi^{\und{a}\und{k}}\,. \lb{newchi}
\ee
Using the supersymmetry transformation laws \p{compN4}, it is  easy to find how ${\cal N}=4$ supersymmetry acts on the $\tau$ basis
bosonic coordinates
\be
\delta f^{ia} = \varepsilon^0\chi^{ia} + \varepsilon^{(\und{l}\und{k})}
\left[e^{\,ia}_{\,\und{l}\und{b}}e^{\,\und{t}\und{b}}_{\,j c}\epsilon_{\und{k}\und{t}}\right]\chi^{jc}\,,\lb{CS1}
\ee
where we split $\varepsilon^{\und{l}\und{k}}$ into singlet and triplet parts:
$$
\varepsilon^{\und{l}\und{k}} = \varepsilon^0 \epsilon^{\und{l}\und{k}} +\varepsilon^{(\und{l}\und{k})}\,.
$$
By the standard reasoning (see e.g. \cite{CoPa}), the coefficient of the triplet part is just the triplet of the complex structures
\be
ie^{\,ia}_{\,(\und{l}\und{b}}e^{\,\und{t}\und{b}}_{\,j c}\epsilon_{\und{k})\und{t}} = I_{(\und{l}\und{k})}{}^{ia}_{jc}\,.\lb{CStr}
\ee
They satisfy the quaternion algebra
\be
I^mI^n = -\delta^{mn} +\epsilon^{mns}I^s\,, \quad I^s \equiv (\sigma^s)^{(\und{l}\und{k})}I_{(\und{l}\und{k})} \lb{Quat}
\ee
and in fact coincide with those in the case of general heterotic $(4,0)$ $d=2$ sigma models \cite{dks}. Now we should check
whether they are covariantly
constant with respect to the generalized affine connection encoded in the Lagrangian \p{Lagr1} - \p{Lagr3}.

\subsection{Metric and connection}
The $\tau$ world metric can be easily read off from the bosonic part of \p{Lagr1}, making use of the relations \p{Viel}-\p{Viel3}:
\be
g_{ia\;kb} = G_{\,[\und{c}\, \und{d}]}\,\epsilon{}_{\,\und{l}\und{t}}\, e^{\,\und{l}\und{c}}_{\,ia}\, e^{\,\und{t}\und{d}}_{\,kb}\,,
\quad g^{ia\;kb} = G^{\,[\und{c}\, \und{d}]}\,\epsilon{}^{\,\und{l}\und{t}}\, e_{\,\und{l}\und{c}}^{\,ia}\, e_{\,\und{t}\und{d}}^{\,kb}\,, \lb{Metr}
\ee
where
\be
 G_{\,[\und{c}\, \und{d}]} = \int du \,{\cal F}_{\,[\und{c}\, \und{d}]}\,, \;\;{\cal F}_{\,[\und{c}\, \und{d}]}
 = {\cal F}_{\,[{c}\, {d}]}\,(M^{-1}){}^c_{\und{c}}\,(M^{-1}){}^d_{\und{d}}\,, \quad
 G_{\,[\und{c}\, \und{d}]}\, G^{\,[\und{d}\, \und{b}]} = \delta^{\,\und{b}}_{\,\und{c}}\,. \lb{DefGund}
\ee

The connection can be extracted by considering that part of the Lagrangian which is bilinear in the fermionic fields.
One should pass to the harmonic-independent fermionic fields $\chi^{\und a}, \bar\chi^{\und{a}}$ by the formulas
\p{Hindchi} and determine the appropriate part of the full equation of motion for fermionic fields (ignoring the tensorial part
coming from the quartic fermionic term \p{Lagr3}) by varying with respect to these harmonic-independent fields. In this way one obtains
\bea
{\cal D}_t\,\bar{\chi}{}^{\,\und{a}} = \dot{\bar{\chi}}{}^{\,\und{a}} &-& G^{\,[\und{a}\,\und{c}]}
\int du \left\{{\cal F}_{\,[\und{c}\,\und{b}]}
\left[\dot{f}^{+\und{d}}\left({\cal D}_{+\und{d}}\,M^{\und{b}}_d\,M^{-1}{}^d_{\und{e}}\right)
+ \nabla f^{-\und{d}}\left({\partial}_{-\und{d}}\,M^{\und{b}}_d\,M^{-1}{}^d_{\und{e}}\right)\right] \right. \nn
&&\left. -\dot{f}^{+\und{b}}\,\tilde{\nabla}_{+\und{e}}\,{\cal F}_{\,[\und{c}\,\und{b}]}
- \nabla f^{-\und{b}}\,\tilde{\nabla}_{-\und{e}}\,{\cal F}_{\,[\und{c}\,\und{b}]}\right\}\bar{\chi}^{\,\und{e}} +
\mbox{\it terms of 3d order in $\chi, \bar\chi$}. \lb{eqmot}
\eea
Here,
\be
\dot{f}^{+\und{d}} = \dot{f}^{+{d}}\,M^{\und{d}}_d\,, \quad \nabla f^{-\und{d}} = \left(\dot{f}^{-d}
- E^{-2 d}_b\dot{f}^{+ b}\right)M^{\und{d}}_d\,,
\ee
and $\tilde{\nabla}_{\pm \und{a}}$ are the covariant derivatives with respect to the harmonic-independent tangent space group acting
on the underlined indices
\bea
\tilde{\nabla}_{+ \und{c}}\,{\cal F}_{\,[\und{a}\,\und{b}]} &=& \nabla_{+ \und{c}}\,{\cal F}_{\,[\und{a}\,\und{b}]}
+ \left({\cal D}_{+\und{c}}\,M^{\und{e}}_d\,M^{-1}{}^d_{\und{a}}\right){\cal F}_{\,[\und{e}\,\und{b}]} +
\left({\cal D}_{+\und{c}}\,M^{\und{e}}_d\,M^{-1}{}^d_{\und{b}}\right){\cal F}_{\,[\und{a}\,\und{e}]} \nn
&=& (M^{-1})^c_{\und c}(M^{-1})^a_{\und a}(M^{-1})^b_{\und b}\;{\cal D}_{+ c}\,{\cal F}_{[ab]} \,, \nn
\tilde{\nabla}_{- \und{c}}\,{\cal F}_{\,[\und{a}\,\und{b}]} &=& \partial_{- \und{c}}\,{\cal F}_{\,[\und{a}\,\und{b}]}
+ \left({\partial}_{-\und{c}}\,M^{\und{e}}_d\,M^{-1}{}^d_{\und{a}}\right){\cal F}_{\,[\und{e}\,\und{b}]} +
\left({\partial}_{-\und{c}}\,M^{\und{e}}_d\,M^{-1}{}^d_{\und{b}}\right){\cal F}_{\,[\und{a}\,\und{e}]} \nn
&=& (M^{-1})^c_{\und c}(M^{-1})^a_{\und a}(M^{-1})^b_{\und b}\;{\partial}_{- c}\,{\cal F}_{[ab]}\,,  \lb{tildenabla}
\eea
where $\nabla_{+\und{c}} = (M^{-1})_{\und{c}}^b\,\nabla_{+ b} = E^{ia}_{+\und{c}}\partial_{ia}\,, \;
\partial_{-\und{c}} = (M^{-1})_{\und{c}}^b\,\partial_{- b} = E^{ia}_{-\und{c}}\partial_{ia}\,$.

Taking into account the fact that the tangent $SU(2)$ indices $\und{i}, \und{k}$ are transformed by the rigid group, the covariant
time derivative of any tangent space vector $P^{\,\und{i}\und{a}}$ will have the same form. Then one can pass to the world vector
$P^{\,ia}$ as
$$
P^{\,ia} = e^{\,ia}_{\,\und{k}\und{b}}\,P^{\,\und{k}\und{b}}
$$
and define the generalized affine connection $\hat{\Gamma}{}^{\,ia}_{jb\;\; lc}$ from the relation
\be
{\cal D}_t\, P^{\,ia} = \dot{f}^{\,kb}\,{\cal D}_{kb}\,P^{\,ia} = \dot{f}^{kb}\left(\partial_{kb}\,P^{\,ia}
+ \hat{\Gamma}{}^{\,ia}_{\,kb \;\; lc}\,P^{\,lc}\right).
\ee
After some algebra we obtain
\bea
\hat{\Gamma}{}^{\,ia}_{\,kb \;\; lc} = e^{\,ia}_{\,\und{t}\und{d}}\,\partial_{\,kb}\,e^{\,\und{t}\und{d}}_{\,lc} +
e^{\,\und{j}\und{d}}_{\,kb}\,e^{\,\und{t}\und{c}}_{\,lc}\,e^{\,ia}_{\,\und{t}\und{a}}\,G^{\,[\und{b}\,\und{a}]}\,
J_{\,\und{j}\,\und{b}\,\und{d}\,\und{c}}\,,\lb{connect1}
\eea
where
\bea
J_{\,\und{j}\,\und{b}\,\und{d}\,\und{c}} &=& \int du \left\{{\cal F}_{\,[\und{e}\,\und{b}]}
\left[u^+_{\und{j}}\left({\cal D}_{+\und{d}}\,M^{\und{e}}_d\,M^{-1}{}^d_{\und{c}}\right)
+ u^-_{\und{j}}\left({\partial}_{-\und{d}}\,M^{\und{e}}_d\,M^{-1}{}^d_{\und{c}}\right)\right] \right. \nn
&&\left. - \,u^+_{\und{j}}\,\tilde{\nabla}{}_{+ \und{c}}\,{\cal F}_{\,[\und{d}\,\und{b}]} - u^-_{\und{j}}\,
 \tilde{\nabla}{}_{- \und{c}}\,{\cal F}_{\,[\und{d}\,\und{b}]}\right\}.\lb{J1a}
\eea
In a more detailed form, the last integral is as follows
\bea
J_{\,\und{j}\,\und{b}\,\und{d}\,\und{c}} &=& \int du \left\{{\cal F}_{\,[\und{e}\,\und{b}]}
\left[u^+_{\und{j}}\,A^{-\,\und{e}}_{\,\und{d}\;\;\und{c}} + u^-_{\und{j}}\,B^{+\,\und{e}}_{\,\und{d}\;\;\und{c}}\right]
-{\cal F}_{\,[\und{d}\,\und{e}]}\left[u^+_{\und{j}}\,C^{-\,\und{e}}_{\,\und{c}\;\;\und{b}}
+ u^-_{\und{j}}\,D^{+\,\und{e}}_{\,\und{c}\;\;\und{b}}\right]\right.\nn
&&\left. +\, e^{\,lb}_{\,\und{j}\und{c}}\,\partial_{\,lb}\,{\cal F}_{\,[\und{d}\,\und{b}]}\right\},\lb{J1}
\eea
where
\bea
&& A^{-\,\und{e}}_{\,\und{d}\;\;\und{c}} = 2\,\nabla_{+[\und{d}}\,M^{\,\und{e}}_d\,M^{-1}{}^d_{\,\und{c}]}
= 2\, C^{-\,\und{e}}_{\,[\und{d}\;\;\und{c}]}\,, \quad
B^{+\,\und{e}}_{\,\und{d}\;\;\und{c}} = 2\,\partial_{-[\und{d}}\,M^{\,\und{e}}_d\,M^{-1}{}^d_{\,\und{c}]}
= 2\, D^{+\,\und{e}}_{\,[\und{d}\;\;\und{c}]}\,, \nn
&& C^{-\,\und{e}}_{\,\und{c}\;\;\und{b}} = E^{-\und{e}}_{\,\und{c}\und{b}} + \nabla_{+\und{c}}\,M^{\,\und{e}}_d\,M^{-1}{}^d_{\,\und{b}}
= {\cal D}_{+\und{c}}\,M^{\,\und{e}}_d\,M^{-1}{}^d_{\,\und{b}}\,, \quad
D^{+\,\und{e}}_{\,\und{c}\;\;\und{b}} = \partial_{-\und{c}}\,M^{\,\und{e}}_d\,M^{-1}{}^d_{\,\und{b}}\,. \lb{ABCD}
\eea

Using some previously derived identities rewritten in the $\tau$ frame, e.g.,
\be
\partial^{++}\,E^{-\und{a}}_{\und{b}\und{c}} = E^{+\und{a}}_{\und{b}\und{c}}\,,
\ee
and also the general matrix identity
\be
d_1\left(A^{-1}\,d_2\,A\right) = - A^{-1}\left[d_2\left(A\,d_1\,A^{-1}\right) \right]A + A^{-1}[d_1, d_2]A
\ee
(which is valid for any bosonic derivation operators $d_1$ and $d_2$), one derives the following relations
\be
\partial^{++}\,D^{+\,\und{e}}_{\,\und{c}\;\;\und{b}} =0\,, \quad \partial^{++}\,C^{-\,\und{e}}_{\,\und{c}\;\;\und{b}} =
- D^{+\,\und{e}}_{\,\und{c}\;\;\und{b}}\,, \lb{RelDC}
\ee
from which it follows that the expressions in the square brackets in \p{J1} do not depend on harmonics. In other words,
the dependence on the extra scalar potential $\tilde{{\cal L}}(f, u)$ manifests itself in the $\tau$ world metrics and connection solely
through the ``symplectic metric'' $G_{\,[\und{a}\,\und{b}]}$ and its inverse. This can be made explicit by noting that \p{RelDC} implies
\be
D^{+\,\und{e}}_{\,\und{c}\;\;\und{b}} = D^{\und{i}\,\und{e}}_{\,\und{c}\;\;\und{b}}u^+_{\und{i}}\,, \quad
C^{-\,\und{e}}_{\,\und{c}\;\;\und{b}} = -D^{\und{i}\,\und{e}}_{\,\und{c}\;\;\und{b}}u^-_{\und{i}}\,. \lb{DefDC}
\ee
{}Using eqs. \p{ABCD}, one can represent $D^{\und{i}\,\und{e}}_{\,\und{c}\,\und{b}}$ as
\be
D_{\und{k}\,\und{d} \;\;\und{a}}^{\;\;\;\und{c}} = e_{\und{k}\und{d}}^{l b}\,\partial_{lb}M^{\und{c}}_{g} \,(M^{-1})^g_{\und{a}}
- u^+_{\und{k}}\,E^{-\und{c}}_{\und{d}\und{a}}\,.
\ee

Substituting \p{DefDC} into \p{J1} yields the final expression for $J_{\,\und{j}\,\und{b}\,\und{d}\,\und{c}} $:
\be
J_{\,\und{j}\,\und{b}\,\und{d}\,\und{c}} = \epsilon_{\und{j}\und{k}}\left(G_{\,[\und{d}\,\und{e}]}\,D^{\und{k}\,\und{e}}_{\,\und{c}\;\;\und{b}}
+2G_{\,[\und{b}\,\und{e}]}\,D^{\und{k}\,\und{e}}_{\,[\und{d}\;\;\und{c}]}\right)
+ e^{\,md}_{\,\und{j}\und{c}}\,\partial_{\,md}\,G_{\,[\und{d}\,\und{b}]}\,. \lb{Jdef}
\ee

The final expression for the connection introduced in \p{connect1} is as follows
\bea
\hat{\Gamma}{}^{\,ia}_{\,kb \;\; lc} &=& e^{\,ia}_{\,\und{t}\und{d}}\,\partial_{\,kb}\,e^{\,\und{t}\und{d}}_{\,lc}
- 2 e^{\,ia}_{\,\und{t}\und{a}}\,e_{\,kb}^{\,\und{j}\und{d}}\,e_{\,lc}^{\,\und{t}\und{c}}\,D^{\und{k}\,\und{a}}_{\,[\und{d}\;\;\und{c}]}
\epsilon_{\und{j}\und{k}}+e^{\,ia}_{\,\und{t}\und{a}}\,e_{\,kb}^{\,\und{j}\und{d}}\,e_{\,lc}^{\,\und{t}\und{c}}\,
G_{\,[\und{d}\,\und{e}]}\,D^{\und{k}\,\und{e}}_{\,\und{c}\;\;\und{b}}\,G^{\,[\und{b}\,\und{a}]}\epsilon_{\und{j}\und{k}} \nn
&& +\, e^{\,\und{j}\und{d}}_{\,kb}\,e^{\,\und{t}\und{c}}_{\,lc}\,e^{\,ia}_{\,\und{t}\und{a}}\,e^{\,md}_{\,\und{j}\und{c}}\,
G^{\,[\und{a}\,\und{b}]}\,\partial_{\,md}\,G_{\,[\und{b}\,\und{d}]}\,. \lb{hatconn}
\eea

It remains to compute
\be
{\cal D}_{ia}\,I_{(\und{k}\und{l})}{}^{\,jb}_{\,td} = \partial_{ia}\,I_{(\und{k}\und{l})}{}^{\,jb}_{\,td} +
\hat{\Gamma}{}^{\,jb}_{\,ia \;\; lc}\,I_{(\und{k}\und{l})}{}^{\,lc}_{\,td}
- \hat{\Gamma}{}^{\,lc}_{\,ia \;\; td}\,I_{(\und{k}\und{l})}{}^{\,jb}_{\,lc}\,,
\ee
where the triplet of complex structures $I_{(\und{k}\und{l})}$ was defined in \p{CStr}.
The direct computation yields zero for this covariant derivative,
\be
{\cal D}_{ia}\,I_{(\und{k}\und{l})}{}^{\,jb}_{\,td} = 0\,,
\ee
i.e. in the general case the triplet of complex structures is {\it also covariantly constant}.
As we shall see soon, this means that the general case with one sort of the multiplets $\bf{(4, 4, 0)}$ yields the weak HKT geometry
(it becomes strong HKT under some additional conditions, see Sect. 5.6).
The most general case, when the covariant constancy condition weakens to
\be
{\cal D}_{(M}\,\tilde{I}^{m}{}^{\,K}_{\;N)} = 0\,,
\ee
should then correspond to the situation when the mirror $\bf{(4, 4, 0)}$ multiplets are also added. We shall consider
this general situation in Sect. 7 in the ${\cal N}=2$ superfield formulation.

\subsection{Torsion}
Let us now compute the torsion. As usual, it is defined as
\be
\hat{\Gamma}_{ia,\; kb\; lc} = \Gamma_{ia,\; kb\; lc} + \frac{1}{2}\,C_{ia \;kb \;lc}\,, \lb{defC}
\ee
where $\Gamma_{ia,\; kb \;lc}$ is the Christoffel symbol for the metric $g_{ia \;lc}$:
\be
\Gamma_{ia,\; kb\; lc} = \frac{1}{2}\,\left(\partial_{kb}\,g_{ia \;lc} + \partial_{lc}\,g_{ia \;kb} - \partial_{ia}\,g_{kb\; lc}\right).
\ee
We represent $\hat{\Gamma}_{ia, \;kb\; lc} = g_{ia\; i'a'}\hat{\Gamma}{}^{\,i'a'}_{\,kb\; lc}$ as
\be
\hat{\Gamma}_{ia, \;kb\; lc} = G_{[\und{c}\, \und{d} ]} \epsilon_{\und{l}\und{t}}\left\{e^{\und{l}\und{c}}_{ia}\;
\partial_{(kb}\,e^{\und{t}\und{d}}_{lc)}
+ e^{\und{l}\und{c}}_{ia}\;\partial_{[kb}\,e^{\und{t}\und{d}}_{lc]} \right\}-
\epsilon_{\und{l}\und{t}}\left(e^{\und{l}\und{u}}_{ia}\, e^{\und{j}\und{h}}_{kb}\,
e^{\und{t}\und{g}}_{lc}\right)J_{\,\und{j}\,\und{u}\,\und{h}\,\und{g}}\,, \lb{GJ}
\ee
where the quantity $J_{\,\und{j}\,\und{u}\,\und{h}\,\und{g}}$ was defined in \p{Jdef}. Using the identities
\be
\partial_{[kb}\,e^{\und{i}\und{d}}_{lc]} = \epsilon_{\und{j}\und{l}}\,e^{\und{i}\und{s}}_{[lc}\,e^{\und{j}\und{b}}_{kb]}\,
D^{\und{l}\und{d}}_{[\und{b}\;\;\und{s}]} - \frac{1}{2}\,\epsilon_{\und{t}\und{j}}\,e^{\und{t}\und{b}}_{[lc}\,
e^{\und{j}\und{s}}_{kb]}D^{\und{i}\und{d}}_{(\und{b}\;\;\und{s})} \lb{Iden1}
\ee
and
\bea
e^{kd}_{\und{j}\und{g}}\partial_{kd}\,G_{[\und{h}\,\und{u}]} + \mbox{cycle}\,(\,\und{g},\,\und{h},\,\und{u}\,) &=& -2\epsilon_{\und{j}\und{k}}
\left\{D^{\und{k}\und{b}}_{[\und{g} \;\;\und{h}]}\, G_{[\und{b}\,\und{u}]} + \mbox{cycle}\,(\,\und{g},\,\und{h},\,\und{u}\,)  \right\}, \lb{Iden2} \\
\partial_{-\und{d}} D^{+\und{a}}_{[\und{c}\;\;\und{b}]} + \mbox{cycle}\,(\,\und{d},\,\und{c},\,\und{b}\,) &=&
-2\left\{ D^{+\und{g}}_{[\und{d}\;\;\und{c}]}\,
D^{+\und{a}}_{[\und{g}\;\;\und{b}]} +
\mbox{cycle}\,(\,\und{d},\,\und{c},\,\und{b}\,)\right\}\,,
\lb{Iden3}
\eea
we were able to show that the torsion
$C_{ia\;kb\;lc}$ is totally antisymmetric, as should be,
\be
C_{ia\;kb\;lc} = C_{[ia\;kb\;lc]}\,,
\ee
and is expressed as
\be
C_{ia\;kb\;lc} =
\epsilon_{\und{l}\und{t}}\,\epsilon_{\und{k}\und{j}}\,
e^{\und{k}\und{s}}_{[ia}\,e^{\und{j}\und{g}}_{kb}\,e^{\und{t}\und{d}}_{lc]}\left\{3
\, \,G_{[\und{c}\,\und{d}]}\, D^{\und{l}\und{c}}_{\und{g}\;\;\und{s}}
+ 2\,G_{[\und{c}\,\und{g}]}\,
D^{\und{l}\und{c}}_{[\und{d}\;\;\und{s}]}\right\} -
2i\,I_{(\und{j}\,\und{t})\,[ia}^{\;\;md}\,
e^{\und{t}\und{h}}_{kb}\,e^{\und{j}\und{g}}_{lc]}\,
\partial_{md}\,G_{[\und{h}\,\und{g}]}\,. \lb{tor0}
\ee

Now, passing to the tangent space indices, after some work the torsion can be represented in the following nice form
\be
C_{\und{i}\und{a}\;\und{k}\und{b}\;\und{l}\und{c}} = e^{ia}_{\und{i}\und{a}} e^{kb}_{\und{k}\und{b}}e^{lc}_{\und{l}\und{c}}\,
C_{{i}{a}\; {k}{b}\;{l}{c}} = \epsilon_{\und{i}\und{l}}\,\nabla_{\und{k}\und{a}} G_{[\und{c}\,\und{b}]}
+ \epsilon_{\und{k}\und{l}}\,\nabla_{\und{i}\und{b}} G_{[\und{a}\,\und{c}]}\,, \lb{tor1}
\ee
where
\be
\nabla_{\und{k}\und{a}} G_{[\und{c}\,\und{b}]} = e^{ka}_{\und{k}\und{a}}\,\partial_{ka}G_{[\und{c}\,\und{b}]}
+ D_{\und{k}\,\und{a}\;\;\und{c}}^{\;\;\;\und{b}'}\,G_{[\und{b}'\,\und{b}]}
+ D_{\und{k}\,\und{a}\;\;\und{b}}^{\;\;\;\und{c}'}\,G_{[\und{c}\,\und{c}']}\,.
\ee
While deriving \p{tor1} from \p{tor0}, we made use of the cyclic identity
\be
\nabla_{\und{k}\und{a}} G_{[\und{c}\,\und{b}]} + \mbox{cycle}\; (\und{a},\; \und{b}, \;\und{c}) = 0\,,\lb{Cycle3}
\ee
which is just another form of the identity \p{Iden2}. The total antisymmetry of the torsion in this representation
can be easily checked, also using \p{Cycle3}.

Let us define, following \cite{dks}, the harmonic projections of $C_{\und{i}\und{a}\;\und{k}\und{b}\;\und{l}\und{c}}$:
\be
C^{\pm\pm\pm}_{\und{a}\,\,\und{b}\,\,\und{c}}
:= u^{\pm \und{i}}u^{\pm \und{k}}u^{\pm \und{l}}C_{\und{i}\und{a}\;\und{k}\und{b}\;\und{l}\und{c}}
\ee
One easily finds
\be
C^{+ + +}_{\und{a}\,\,\und{b}\,\,\und{c}} = 0\,, \lb{11}
\ee
and, using \p{Cycle3},
\be
C^{++-}_{\und{a}\,\,\und{b}\,\,\und{c}} = u^{+\und{k}}\nabla_{\und{k}\und{c}} G_{[\und{a}\,\und{b}]}\,, \quad
C^{+-+}_{\und{a}\,\,\und{b}\,\,\und{c}} = u^{+\und{k}}\nabla_{\und{k}\und{b}} G_{[\und{c}\,\und{a}]}\,, \quad
C^{-++}_{\und{a}\,\,\und{b}\,\,\und{c}} = u^{+\und{k}}\nabla_{\und{k}\und{a}} G_{[\und{b}\,\und{c}]}\,. \lb{12}
\ee
Then it is easy to show that
\be
C^{++-}_{[\und{a}\,\,\und{b}\,\,\und{c}]} = 0\,, \quad C^{++-}_{\und{a}\,\,\und{b}\,\,\und{c}} = -C^{+-+}_{\und{a}\,\,\und{c}\,\,\und{b}} =
-C^{++-}_{\und{b}\,\,\und{a}\,\,\und{c}} = C^{-++}_{\und{c}\,\,\und{a}\,\,\und{b}}\,.\lb{13}
\ee
The conditions \p{11} and \p{13} exactly coincide with the harmonic projections of the torsion constraints derived in \cite{dks}. These
constraints, without further closedness restriction on the torsion, define what is now called {\it weak} HKT geometry.
Thus the general system of self-interacting $({\bf 4, 4, 0)}$ ${\cal N}=4, d=1$ multiplets  reveals the most general weak HKT geometry
as its target geometry.

Finally, note a number of useful relations and identities. First, the identity \p{Iden1} allows one to express the object
$D^{\und{i}\,\und{d}}_{\und{b}\,\und{c}}$ in terms of vielbeins:
\be
D^{\und{i}\,\und{d}}_{\und{f}\;\;\und{g}} = \frac{2}{3}\,e^{k\,b}_{[\und{t}\,\und{f}}\,e^{l\,c}_{\und{j}\,\und{g}]}
\left(\epsilon^{\und{j}\und{t}}\,\partial_{k\,b} e^{\und{i}\,\und{d}}_{l\,c} +  \epsilon^{\und{i}\und{t}}\,\partial_{k\,b}
e^{\und{j}\,\und{d}}_{l\,c}\right). \lb{Iden5}
\ee
Also note that \p{Iden3} is a consequence of the following more general identity
\bea
\partial_{-\und{d}} D^{+\und{a}}_{\und{c}\;\;\und{b}} - \partial_{-\und{c}} D^{+\und{a}}_{\und{d}\;\;\und{b}} =
2\,D^{+ \und{a}}_{\und{e}\;\;\und{b}}D^{+ \und{e}}_{[\und{c}\;\;\und{d}]} + D^{+ \und{e}}_{\und{c}\;\;\und{b}}D^{+\und{a}}_{\und{d}\;\;\und{e}}
 - D^{+ \und{e}}_{\und{d}\;\;\und{b}}D^{+ \und{a}}_{\und{c}\;\;\und{e}}\,, \lb{Iden4}
\eea
which is equivalent to the relation
\bea
e_{(\und{k}\und{d}}^{ic}\,\partial_{ic} D^{\;\;\;\und{a}}_{\und{l})\,\und{c}\;\;\und{b}}
-  e_{(\und{k}\und{c}}^{ic}\,\partial_{ic}D^{\;\;\;\und{a}}_{\und{l})\,\und{d}\;\;\und{b}} =
2\,D^{\;\;\;\und{a}}_{(\und{k}\,\und{e}\;\;\und{b}}D^{\;\;\;\und{e}}_{\und{l})\,[\und{c}\;\;\und{d}]}
+ D^{\;\;\;\und{e}}_{(\und{k}\,\und{c}\;\;\und{b}}D^{\;\;\;\und{a}}_{\und{l})\,\und{d}\;\;\und{e}}
 - D^{\;\;\;\und{e}}_{(\und{k}\,\und{d}\;\;\und{b}}D^{\;\;\; \und{a}}_{\und{l})\,\und{c}\;\;\und{e}}\,. \lb{Iden5a}
\eea
These relations ensure the $\tau$ frame form of the conditions for preservation of the target space analyticity
$$
[\tilde{\nabla}_{-\und{a}}, \tilde{\nabla}_{-\und{b}}] = [\tilde{\nabla}_{+\und{a}}, \tilde{\nabla}_{+\und{b}}] = 0
$$
(their $\lambda$ frame form is given by the first two relations in \p{CommA}).

Two other helpful identities are corollaries of \p{Iden1}:
\bea
&& e^{ib}_{(\und{j}\und{d}} \,\partial_{ib}e^{lc}_{\und{k})\und{a}} - e^{ib}_{(\und{j}\und{a}} \,\partial_{ib}e^{lc}_{\und{k})\und{d}} =
\epsilon_{\und{j}\,\und{l}}\,e^{lc}_{\und{k}\und{e}}D^{\und{l}\,\und{e}}_{[\und{a}\;\; \und{d}]} +
\epsilon_{\und{k}\,\und{l}}\,e^{lc}_{\und{j}\und{e}}D^{\und{l}\,\und{e}}_{[\und{a}\;\; \und{d}]}\,, \lb{Iden6} \\
&& e^{kb}_{[\und{j}\und{d}}\,\partial_{kb} e^{lc}_{\und{i}\und{a}]}\, e^{\und{t}\und{e}}_{lc} = \frac{1}{2}
\left(\delta^{\und{t}}_{\und{j}}\,D_{\und{i}\,\und{a}\;\;\und{d}}^{\;\;\;\und{e}} -
\delta^{\und{t}}_{\und{i}}\,D_{\und{j}\und{d}\;\;\und{a}}^{\;\;\;\und{e}}\right). \lb{Iden9}
\eea

\subsection{Spin connection}
Let us calculate the spin connection associated with the affine connection $\hat{\Gamma}$, eq. \p{hatconn}. It is defined by the standard formula
\be
(\omega_{ia})^{\;\;\und{j}'\und{b}'}_{\und{j}\und{b}} = e^{\und{j}'\und{b}'}_{{j}{b}}\left[\partial_{ia}e^{{j}{b}}_{\und{j}\und{b}}
+ \hat{\Gamma}^{{j}{b}}_{{i}{a}\; kd}\,e^{{k}{d}}_{\und{j}\und{b}}  \right]. \lb{DefOm}
\ee
It is straightforward to find
\bea
(\omega_{ia})^{\;\;\und{j}'\und{b}'}_{\und{j}\und{b}} = \delta^{\und{j}'}_{\und{j}}\, e^{\und{i}\und{a}}_{ia}\,
(\omega_{\und{i}\und{a}})^{\;\und{b}'}_{\und{b}}\,, \qquad
(\omega_{\und{i}\und{a}})^{\;\und{b}'}_{\und{b}} = - D_{\und{i}\und{a}\;\;\und{b}}^{\;\;\;\und{b}'} + G^{[\und{b}'\,\und{c}]}\,
\nabla_{\und{i}\und{b}} G_{[\und{c}\,\und{a}]}\,. \lb{om1}
\eea

Using the cyclic identity \p{Cycle3}, the ``reduced'' connection $(\omega_{\und{i}\und{a}})^{\;\und{b}'}_{\und{b}}$
can be cast in another suggestive form
\be
(\omega_{\und{i}\und{a}})^{\;\und{b}'}_{\und{b}} = G^{[\und{b}'\,\und{d}]}\left[(\omega_{\und{i}\und{a}})_{(\und{d}\,\und{b})}
+ (\omega_{\und{i}\und{a}})_{[\und{d}\,\und{b}]} \right],
\ee
where
\bea
&&(\omega_{\und{i}\und{a}})_{(\und{d}\,\und{b})} = D_{\und{i} \und{d} \;\;[\und{b} \, \und{a}]} + D_{\und{i} \und{b} \;\;[\und{d} \, \und{a}]} -
D_{\und{i}\und{a}\;\; (\und{d}\,\und{b})} + e^{md}_{\und{i} (\und{d}}\partial_{md}\,G_{[\und{b})\,\und{a}]} \,, \nn
&&
D_{\und{i}\und{a}\;\; \und{d}\,\und{b}}:= G_{[\und{d}\,\und{d}']}D_{\und{i}\und{a}\;\; \und{b}}^{\;\;\;\und{d}'}\,, \lb{symm}
\eea
and
\be
(\omega_{\und{i}\und{a}})_{[\und{d}\,\und{b}]}=
\frac{1}{2}\,e^{k c}_{\und{i}\und{a}}\partial_{kc} G_{[\und{d}\,\und{b}]}\,. \lb{antisymm}
\ee
The $gl(2n)\ominus usp(2n)$ part of the connection \p{antisymm} vanishes in the admissible $\tau$ frame gauge $G_{[\und{b}\,\und{c}]} =
\Omega_{[\und{b}\,\und{c}]}$ (see Subsect. 5.5), leaving us with the $USp(2n)$ connection $(\omega_{ia})_{(db)}$ as the only essential part
of the spin connection in the general case of weak HKT geometry.

We also observe from \p{om1} that the $GL(2n)$ connection $D_{\und{i}\und{a}\;\;\und{b}}^{\;\;\;\und{b}'}$ appearing in the $\tau$-frame
covariant derivative $\nabla_{\und{i}\,\und{a}}$ differs from the ``canonical'' connection $(\omega_{\und{i}\und{a}})^{\;\und{b}'}_{\und{b}}$ by
a tensor which is just the $\tau$-frame covariant derivative of the ``symplectic'' metric $G_{[\und{c}\,\und{a}]}$. If we define a new
$\tau$-frame covariant derivative $\nabla_{\und{i}\,\und{a}}^{(\omega)}\,$, just with the connection $- \omega$ in place of $D$, we
find, using the identity \p{Cycle3}, that
\be
\nabla_{\und{i}\,\und{a}}^{(\omega)}G_{[\und{b}\,\und{c}]} = 0\,, \lb{covconst}
\ee
i.e. $G_{[\und{b}\,\und{c}]}$ is covariantly constant with respect to the covariant derivative with the connection $\omega\,$.
The identity \p{covconst} resembles the standard Riemann postulate for the metric.

Note that the object $J_{\,\und{j}\,\und{b}\,\und{d}\,\und{c}}$ defined in \p{Jdef} actually coincides with the
``reduced'' spin connection defined in \p{om1}:
\be
J_{\,\und{j}\,\und{b}\,\und{d}\,\und{c}} = -G_{[\und{b}\;\und{b}']}(\omega_{\und{j}\,\und{d}})^{\und{b}'}_{\und{c}}\,,
\ee
while the relation \p{GJ} is just another form of \p{DefOm}.

\subsection{Gauge}
In what follows, it will be sometimes convenient to choose the appropriate gauge with respect to the tangent space $\tau$ gauge transformations.
From the natural assumption about the existence of the flat limit of our Lagrangian, the field-dependent skew-symmetric ``metric''
$G_{[\und{c} \und{d}]}$ should have as its flat limit the standard constant $USp(2n)$ metric $\Omega_{[\und{c} \und{d}]}$:
\be
G_{[\und{c} \und{d}]}= \Omega_{[\und{c} \und{d}]} + \tilde{G}_{[\und{c} \und{d}]}\,.
\ee
Then, taking into account that the generic tangent space gauge group transforms $G_{[\und{c} \und{d}]}$ as
\be
\delta G_{[\und{c} \und{d}]} = \lambda_{\und{c}}^{\;\;\und{c}'}G_{[\und{c}' \und{d}]}
+\lambda_{\und{d}}^{\;\;\und{d}'}G_{[\und{c} \und{d}']}\,, \lb{TangTran}
\ee
it is clear that the whole $G_{[\und{c} \und{d}]}$ can be gauged into its flat part by fixing the antisymmetric part
of the parameter $\lambda_{\und{c}}^{\;\;\und{c}'}$:
\be
G_{[\und{c}\, \und{d}]}= \Omega_{[\und{c}\, \und{d}]}\,, \quad G^{[\und{c}\, \und{d}]}= \Omega^{[\und{c}\, \und{d}]}\,. \lb{TangGauge}
\ee
The residual gauge group is $USp(n)$ with the symmetric parameters $\lambda_{(\und{c}\,\und{d})}$:
\be
\lambda_{\und{c}}^{\;\;\und{c}'} \sim \Omega^{[\und{c}'\,\und{d}' ]}\lambda_{(\und{c}\,\und{d}' )}\,.
\ee

Many relations and identities are simplified in this gauge, e.g.,
\bea
&& \nabla_{\und{i}\und{b}} G_{[\und{c}\,\und{a}]} = 2 D_{\und{i}\,\und{b}\;\;[\und{c}\,\und{a}]}\,, \quad
C_{\und{i}\und{a}\;\und{k}\und{b}\;\und{l}\und{c}} = 2\left(\epsilon_{\und{i}\und{l}}\,D_{\und{k}\,\und{a}\;\;[\und{c}\,\und{b}]}
+ \epsilon_{\und{k}\und{l}}\,D_{\und{i}\,\und{b}\;\;[\und{a}\,\und{c}]}\right),\lb{gaugeCOm1} \\
&& (\omega_{\und{i}\und{a}})_{\und{d}\,\und{b}} = (\omega_{\und{i}\und{a}})_{(\und{d}\,\und{b})} =
D_{\und{i} \und{d} \;\;[\und{b} \, \und{a}]} + D_{\und{i} \und{b}\;\; [\und{d} \, \und{a}]} -
D_{\und{i}\und{a} (\und{d}\,\und{b})}\,, \lb{gaugeCOm2}
\eea
etc. We see that in the gauge \p{TangGauge} the reduced spin connection $(\omega_{\und{i}\und{a}})_{\und{d}\,\und{b}}$ becomes
symmetric, i.e. $usp(2n)$ algebra-valued, in agreement with the fact that the residual tangent-space gauge group is just $USp(2n)$.

One can wonder how the presence of the additional potential ${\cal L}(f, u)$ and the associate symplectic metric
${\cal F}_{[c \,d]}$ reveals itself in the gauge \p{TangGauge}. Naively, the possibility to choose this gauge could be regarded
as a signal that the $\tau$ frame geometry (metric, torsion, ...) does not depend on ${\cal F}_{[c \,d]}$ at all and is
fully specified by the analytic potential ${\cal L}^{+ 3 a}(f^+, u)\,$. To see that this assertion is premature, let us recall
the general expressions for the frame bridges \p{Redund}. It is clear that the gauge \p{TangGauge} amounts to fix the $\tau$ gauge
matrices $L^{\und{a}}_{\und{b}}$ in the following way:
\be
L^{\und{a}}_{\und{b}}\,\Omega_{[\und{a}\,\und{c}]}\,L^{\und{c}}_{\und{d}} = \tilde{G}_{[\und{b}\,\und{d}]} = \int du\,
{\cal F}_{\,[{c}\, {d}]}\,(\tilde{M}^{-1}){}^c_{\und{c}}\,(\tilde{M}^{-1}){}^d_{\und{d}}\,, \lb{Gauge5}
\ee
where the bridge $\tilde{M}_c^{\und{b}}$, as defined in \p{Redund}, is fully specified by ${\cal L}^{+ 3 a}(f^+, u)\,$.
Thus in the gauge \p{TangGauge} the metric ${\cal F}_{[c \,d]}$ appears implicitly,
through the non-trivial $\tau$ gauge factor $L^{\und{a}}_{\und{b}}$ in the full bridge ${M}_c^{\und{b}} =
\tilde{M}_c^{\und{d}}\,L^{\und{b}}_{\und{d}}$. This harmonic-independent matrix factor, up to the residual $USp(2n)$ gauge freedom,
is defined by eq. \p{Gauge5}. An alternative possibility would be from the very beginning to fully or partly fix the $\tau$ gauge freedom  by
choosing, e.g., in \p{Redund}, $L^{\und{a}}_{\und{b}} = \delta^{\und{a}}_{\und{b}}$. In such gauges, the symplectic
metric $G_{[\und{c} \,\und{d}]}$ explicitly persists in all formulas of the $\tau$ frame geometry.

The simplest example of this gauge equivalence is provided by the case with ${\cal L}^{+ 3a} = 0\,$ that corresponds
to the linear $({\bf 4,4,0})$ multiplets. In this case
$\tilde{M}{}^{\und{d}}_a = \delta {}^{\und{d}}_a\,, \; (\tilde{M}^{-1})_{\und{d}}^a = \delta_{\und{d}}^a$ and the natural
$\tau$ gauge choice is $L^{\und{d}}_{\und{a}} = \delta^{\und{d}}_{\und{a}}$, which yields
\be
g_{ia\,kb} = G_{[a\,b]}\epsilon_{ik}\,, \quad G_{[a\,b]} = \int du\, {\cal F}_{[a\,b]} =
\frac{1}{2}\Delta_{[a\,b]}\int du \,{\cal L}(f, u) \lb{metr11}
\ee
in agreement with \p{cflatn}. Alternatively, one can choose the gauge \p{TangGauge} which in the considered case amounts to
\be
G_{[\und{c}\,\und{b}]} = \Omega_{[\und{c}\,\und{b}]}\,, \; M{}^{\und{d}}_a = L{}^{\und{d}}_{\und{b}}\,\delta^{\und{b}}_a\,, \quad
L^{\und{a}}_{\und{b}}\,\Omega_{[\und{a}\,\und{c}]}\,L^{\und{c}}_{\und{d}} = \delta_{\und{b}}^a \,\delta_{\und{d}}^c\,G_{[a\,c]}\,.
\ee
After substituting this into the general expression \p{Metr}, one recovers the same final expression \p{metr11} for the $\tau$ frame metric.

\subsection{Torsion closedness conditions}
Let us examine under which conditions the torsion $C_{\und{i}\und{a}\;\und{k}\und{b}\;\und{l}\und{c}}$ is closed. This problem
is most simply treated using the world indices representation, where the closedness condition is just
\be
\partial_{[jd}\,C_{ia\;kb\;lc]} = 0\,. \lb{closed}
\ee

We will use the relation:
\be
\nabla_{\und{i}\,\und{a}}\,G_{[\und{c}\,\und{d}]} = - \int du (M^{-1})_{\und{a}}^{{a}'}(M^{-1})_{\und{c}}^{{c}'}
(M^{-1})_{\und{d}}^{{d}'}\left(\partial_{-a'}\,{\cal F}_{[c'\,d']} u^-_{\und{i}}
+ {\cal D}_{+ a'}{\cal F}_{[c'\,d']} u^+_{\und{i}}\right).\lb{repre1}
\ee
After some work, the world-indices form of \p{tor1} is found to be
\be
C_{ia\;kb\;lc} = \partial_{ia}\,{\cal B}_{kb\;lc} + \partial_{lc}\,{\cal B}_{ia\;kb} + \partial_{kb}\,{\cal B}_{lc\;ia}
+ \hat{C}_{ia\;kb\;lc}\,,\lb{tor3}
\ee
where
\be
{\cal B}_{ia\;kb} = -\int du \,{\cal F}_{[g\,f]}\,\left(E^{+ g}_{ia}E^{-f}_{kb} -  E^{+ g}_{kb}E^{-f}_{ia}\right), \lb{pot}
\ee
and
\be
\hat{C}_{ia\;kb\;lc} = 2 \int du \left[E^{-u}_{ia}\partial_{- u}{\cal F}_{[g\,f]}\left( E^{+ g}_{kb}E^{-f}_{lc} -  E^{+ g}_{lc}E^{-f}_{kb}\right)
+ \mbox{cycle}\,(ia, \; kb,\;lc) \right]. \lb{defChat}
\ee

The representation \p{tor3} is convenient in that it reduces the condition \p{closed} to
\be
\partial_{[jd}\,\hat{C}_{ia\;kb\;lc]} = 0\,, \lb{closed1a}
\ee
which is easier to solve. It proves to be equivalent to the condition
\be
\int du(M^{-1})_{\und{d}}^{{d}}(M^{-1})_{\und{a}}^{{a}} (M^{-1})_{\und{b}}^{{b}}(M^{-1})_{\und{c}}^{{c}}\left({\cal D}_{+ {d}}\partial_{-{a}}
- {\cal D}_{+ {a}}\partial_{-{d}} \right){\cal F}_{[{b}\,{c}]} = 0\,. \lb{closed1}
\ee
We see that this condition amounts to the vanishing of the contribution of the four-fermionic term \p{Lagr3} to the component
harmonic-independent Lagrangian
$$
{L}^{(3)} = \int du \,{\cal L}{}^{(3)} = 0\,.
$$

The same condition \p{closed1}, rewritten in the $\tau$ frame, reads
\be
\epsilon^{\und{i}\und{k}}\,\nabla_{\und{i}[\und{a}} \nabla_{\und{k}\und{b}]}\,
G_{[\und{c}\und{d}]} = 0\,.\lb{close2}
\ee
In the four-dimensional case, the differential operator in \p{close2} becomes a generalization  of the covariant Laplace-Beltrami
operator (see Sect. 6). Note
that the symplectic metric $G_{[\und{c}\und{d}]}$ in the general case can be expressed as
\be
G_{[\und{c}\und{d}]} =
\frac{1}{2}\,\epsilon^{\und{i}\,\und{k}}\,\nabla_{\und{i}[\und{c}} \nabla_{\und{k}\und{d}]}\,
L(f)\,, \qquad L(f):= \int du\, {\cal L}(f, u)\,, \lb{exprG}
\ee
where ${\cal L}(f, u)$ is defined in \p{DefL}. This representation for $G_{[\und{c}\und{d}]}$ is an obvious generalization
of the representation \p{cflatn} which is valid in the case of linear ${\bf (4,4,0)}$ multiplets.

It is worth noting that the condition \p{close2} can be also derived more directly, starting from the $\tau$ frame form of \p{closed},
i.e.
\be
\nabla_{[\und{j}\und{d}}\,C_{\und{i}\und{a}\;\und{k}\und{b}\;\und{l}\und{c}]} = 0\,, \lb{closed4}
\ee
and taking into account the relation $[\nabla_{(\und{j}\und{d}}, \nabla_{\und{k})\und{b}}] = 0\,$, which follows from
\p{Iden5a} and \p{Iden9}. Actually, the vanishing of this symmetrized commutator is the $\tau$ frame form of the target-space
analyticity preservation conditions.

It is also interesting to see how the closedness condition looks in the gauge \p{TangGauge}:
\be
\nabla_{\und{k}\,\und{a}} D^{\und{k}}_{\und{b}\;\;[\und{c}\,\und{d}]}
- \nabla_{\und{k}\,\und{b}} D^{\und{k}}_{\und{a}\;\;[\und{c}\,\und{d}]} = 0\,. \lb{close3}
\ee
This can be treated as an extra condition on the bridges $M_c^{\und{b}}$. After some work, an equivalent condition
in the $\lambda$-frame is found to be
\be
{\cal D}_{+[a}\partial_{-b]}\,T_{[c\,d]} =  \partial_{-[b}{\cal D}_{+a]}\,T_{[c\,d]} = 0\,, \lb{close4}
\ee
where
\be
T_{[c\,d]} = M^{\und{a}}_c\,M^{\und{b}}_d\,\Omega_{\und{a}\,\und{b}}\,. \lb{defT}
\ee
Note that only the $GL(2n)/USp(2n)$ coset part of the bridge actually contributes to $T_{[c\,d]}\,$.

\subsection{The $\tau$ world Lagrangians}
Here we rewrite the component Lagrangians in the invariant actions \p{nonlqAct} and \p{WZ2}
in terms of the objects of the $\tau$ world geometry. Passing to the $\tau$ frame form of the actions can be
accomplished with the help of bridges, like in Chapter 11.4.2 of \cite{HSS1}.

We define the component sigma-model Lagrangian as
\be
{L}^{comp}(f^{ia}, \chi^{\und{a}}, \bar\chi{}^{\und{b}}) = -\frac{1}{8}\int du \,{\cal L}^{comp}(f^\pm, \chi, \bar\chi, u^\pm)\,,
\ee
where ${\cal L}^{comp}$ was defined in eqs. \p{nonlqAct2} - \p{DefL}. After some work, using the relations of this Section,
the Lagrangian
${L}^{comp}$ can be written in the following simple form
\be
{L}^{comp} = \frac{1}{2}\,g_{ia\, kb}\,\dot{f}^{ia}\dot{f}^{kb}
-\frac{i}{4}\,G_{[\und{a}\,\und{b}]}\left(\nabla \bar{\chi}^{\und{a}}\chi^{\und{b}}
- \bar{\chi}^{\und{a}}\nabla \chi^{\und{b}}\right)
- \frac{1}{16}\left(\epsilon^{\und{i}\,\und{k}}\nabla_{\und{i}[\und{a}}\nabla_{\und{k}\und{b}]}
\,G_{[\und{c}\,\und{d}]}\right)\bar\chi{}^{\und{a}}\bar\chi{}^{\und{b}}\chi^{\und{c}}\chi^{\und{d}}\,.\lb{taulagr1}
\ee
Here
\be
\nabla \chi^{\und{b}} = \dot{\chi}{}^{\und{b}} + \dot{f}{}^{kb} e^{\und{i}\und{a}}_{kb}\,(\omega_{\und{i}\und{a}})^{\und{b}}_{\;\;\und{d}}\,
\chi^{\und{d}}\,, \quad \nabla \bar\chi^{\und{b}} = \dot{\bar\chi}{}^{\und{b}} + \dot{f}{}^{kb} e^{\und{i}\und{a}}_{kb}\,
(\omega_{\und{i}\und{a}})^{\und{b}}_{\;\;\und{d}}\,\bar\chi^{\und{d}}\,,
\ee
and $(\omega_{\und{i}\und{a}})^{\und{b}}_{\;\;\und{d}}$ is the ``reduced'' spin connection defined in \p{DefOm}, \p{om1}. Note that
the four-fermionic term in \p{taulagr1} can be rewritten in a more geometric way through the external derivative of the torsion tensor as
\be
\frac{1}{16}\left(\epsilon^{\und{i}\,\und{k}}\nabla_{\und{i}[\und{a}}\nabla_{\und{k}\und{b}]}
\,G_{[\und{c}\,\und{d}]}\right)\bar\chi{}^{\und{a}}\bar\chi{}^{\und{b}}\chi^{\und{c}}\chi^{\und{d}} =
\frac{1}{4!}\epsilon^{\und{j}\,\und{k}}\,\epsilon^{\und{i}\,\und{l}}\,\nabla_{[\und{j}\und{a}} \,
C_{\und{i}\und{b}\;\und{k}\und{c}\;\und{l}\und{d}]}\,\bar\chi{}^{\und{a}}\bar\chi{}^{\und{b}}\chi^{\und{c}}\chi^{\und{d}}\,.
\ee
In this notation, there becomes clear, e.g., the invariance of the coefficient with respect to the permutation
$[\und{a}\,\und{b}] \leftrightarrow [\und{c}\,\und{d}]\,$. Also, it is immediately seen that the four-fermionic term vanishes
under the closedness torsion condition \p{closed4} (and, of course, in the torsionless HK case).

The Wess-Zumino component Lagrangian
\be
{L}_{WZ} = -\frac{1}{2}\int du \,{\cal L}_{WZ}\,,
\ee
where ${\cal L}_{WZ}$ was defined in \p{WZcomp}, after integration over harmonics can be represented as
\be
{L}_{WZ} = {\cal A}_{ia}(f)\dot{f}^{ia} -\frac{i}{2}\left(\int du\,{\cal D}_{+ \und{a}}\nabla_{+ \und{b}}{\cal L}^{+ 2}\right)\,
\chi^{(\und{a}}\bar\chi^{\und{b})}\,,\lb{WZcomp11}
\ee
where
\be
{\cal A}_{ia}(f) = e_{ia}^{\und{k}\und{b}}\, \int du\,u^-_{\und{k}}\,\nabla_{+\und{b}}{\cal L}^{+ 2}(f^+, u)\,.
\ee
The target analytic gauge transformations \p{GaugeCom} produce the standard abelian gauge transformation of ${\cal A}_{ia}(f)$:
\be
{\cal A}_{ia}(f) \;\rightarrow \; {\cal A}_{ia}(f) - \partial_{ia}\sigma(f)\,, \quad \sigma(f) = \int du \,\sigma(f^+, u)\,.
\ee
Defining
\be
{\cal F}_{\und{i}\und{a}\;\und{k}\und{b}} = e^{jd}_{\und{i}\und{a}}\, e^{l c}_{\und{k}\und{b}}\,{\cal F}_{jd\;lc}\,, \quad
{\cal F}_{jd\;lc} = \partial_{jd}\,{\cal A}_{lc} - \partial_{lc}\,{\cal A}_{jd}\,,
\ee
it is easy to find
\be
{\cal F}_{\und{i}\und{a}\;\und{k}\und{b}} = -\epsilon_{\und{i}\,\und{k}}\,\int du\,{\cal D}_{+ (\und{a}}\nabla_{+ \und{b})}{\cal L}^{+ 2}\,.
\ee
This implies that the background gauge field ${\cal A}_{ia}(f)$ in the general case is self-dual,
${\cal F}_{(\und{i}\und{a}\;\und{k})\und{b}} = 0\,$, like in the case of linear $({\bf 4, 4, 0})$ multiplets \cite{IL,ikons}.
Thus the self-duality of the background gauge field with respect to the $\tau$ frame $SU(2)$ doublet  indices is a general necessary
condition of implementation of ${\cal N}=4$ supersymmetry in the WZ term of the $(\bf{4, 4, 0})$ multiplets. The Lagrangian
\p{WZcomp11} can be rewritten as
\be
{L}_{WZ} = {\cal A}_{ia}\,\dot{f}^{ia} -\frac{i}{4}\,\epsilon^{\und{i}\,\und{k}}\,
{\cal F}_{\und{i}\und{a}\;\und{k}\und{b}}\,\chi^{\und{a}}\bar\chi^{\und{b}}\,.\lb{WZcomp12}
\ee

\setcounter{equation}{0}
\section{Particular cases}
\subsection{Hyper-K\"ahler case}

The HK case  corresponds to the choice \p{HK}, \p{L4tran}  for the analytic potential ${\cal L}^{+ 3 a}\,$,
\be
{\cal L}^{+ 3 a} = \Omega^{[a\,b]}\partial_{+ b}\,{\cal L}^{+4}(f^+, u)\,, \lb{HK2}
\ee
and the choice
\be
{\cal L}(f, u) = \Omega_{[a\,b]}\,f^{+ a}f^{-b} \quad \Rightarrow \quad {\cal F}_{[a\,b]} =
\partial_{+[a}\partial_{-b]}\,{\cal L}(f, u) =\Omega_{[a\,b]}\,. \lb{constHK}
\ee
Indeed, in this case the basic harmonic constraint \p{Bas} becomes
\be
\partial^{++}f^{+ a} = \Omega^{[a\,b]}\,\partial_{+ b}\,{\cal L}^{+ 4}(f^+, u)\,,
\ee
which is just the constraint defining a general HK sigma model with $4n$-dimensional target space
in the HSS approach \cite{HSS,HSS1,geom1}, with ${\cal L}^{+ 4}(f^+, u)$ being a generic analytic HK potential.
The $\lambda$ world metric extracted from \p{Lagr1} and its $\tau$ world counterpart \p{Metr}, \p{DefGund} in this case coincide
with the expressions derived in \cite{HSS,HSS1,geom1}\footnote{This coincidence holds up to some numerical factors and
a different (though equivalent) choice of the non-analytic $\lambda$-world coordinate $f^{- a}\,$.}.

As follows from the definitions \p{dEf2} and \p{DefEabc}, the constraint \p{HK2} implies
\be
E^{+ 2a}_b = \Omega^{[a d]}E^{+2}_{(db)}\,, \qquad E^{+ a}_{bc} =  \Omega^{[a d]}E^+_{(dbc)}\,, \lb{HKan}
\ee
where
\be
E^{+2}_{(db)} := \partial_{+d}\,\partial_{+b}\,{\cal L}^{+ 4}\,, \qquad E^+_{(dbc)}
:= \partial_{+d}\,\partial_{+b}\,\partial_{+c}\,{\cal L}^{+ 4}\,.
\ee
Then, applying the lemma \p{lemma}, for the non-analytic vielbeins defined by eqs. \p{ZC2} and \p{DerCo2} we obtain the analogous restrictions
\be
E^{- 2a}_b = \Omega^{[a d]}E^{-2}_{(db)}\,, \qquad E^{- a}_{bc} =  \Omega^{[a d]}E^-_{(dbc)}\,.\lb{HKnonan}
\ee

Making use of the constancy of ${\cal F}_{[a\,b]}$ in \p{constHK} and the second relation in \p{HKnonan}, it is easy to find that in the HK case
\be
\partial_{-a}{\cal F}_{[c\,d]} = {\cal D}_{+ a}{\cal F}_{[c\,d]} = 0\,,
\ee
whence, as follows from the representation \p{repre1},
\be
\nabla_{\und{i}\und{a}}\,G_{[\und{c}\,\und{d}]} = 0\,. \lb{zero1}
\ee
This, in turn, implies the vanishing of the torsion defined by \p{tor1}:
\be
C_{ia\;kb\;lc} = C_{\und{i}\und{a}\;\und{k}\und{b}\;\und{l}\und{c}} = 0\,,\lb{zero2}
\ee
as should be in the HK case. Thus the affine connection $\hat{\Gamma}$ coincides with Christoffel symbols $\Gamma\,$.

The bridges defined by \p{FrBrid} are $USp(2n)$-valued in the HK case,
so the connection $D^{\;\;\;\und{e}}_{\und{i}\und{c}\;\;\und{b}}$
defined by \p{ABCD}, \p{DefDC} is symmetric
\be
\Omega_{[\und{e}\,\und{d}]}D^{\;\;\;\und{d}}_{\und{i}\und{c}\;\;\und{b}}
= \Omega_{[\und{b}\,\und{d}]}D^{\;\;\;\und{d}}_{\und{i}\und{c}\;\;\und{e}}\,, \lb{traceless}
\ee
and so is $usp(2n)$-algebra valued. It follows from \p{om1} and \p{zero1} that
\be
(\omega_{\und{i}\und{a}})^{\;\und{b}'}_{\und{b}} = - D_{\und{i}\,\und{a}\;\;\und{b}}^{\;\;\;\und{b}'}\,,
\ee
i.e. we are left with the standard HK $usp(2n)$ spin connection. Both the $\lambda$ and $\tau$ frame gauge transformations are now
$USp(2n)$ ones; the $\lambda$ frame gauge transformations are specified by eqs. \p{spnparam}, \p{Constr} and \p{L4tran}.
Note that the $USp(2n)$ bridges satisfy the relation
$$
\Omega_{[c\,d]}(M^{-1})^c_{\und{c}}(M^{-1})^d_{\und{d}} = \Omega_{[\und{c}\,\und{d}]}\,,
$$
so the $\tau$ frame symplectic metric $G_{[\und{c}\,\und{d}]}$ defined in \p{DefGund} also coincides with its constant part
$$
G_{[\und{c}\,\und{d}]} = \Omega_{[\und{c}\,\und{d}]}\,.
$$

It is worth noting that the superfield action for the {\it general} HK model has the very simple universal form
\be
\tilde{S}_q^{hk} = \int du dt d^4\theta\,\Omega_{[a\,b]}\,q^{+ a}D^{--}q^{+b}
= -2i \int du d\zeta^{(-2)}\,\Omega_{[a\,b]}\,q^{+ a}\partial_t q^{+ b}\,, \lb{ActHK}
\ee
that is, looks the same as the action of the free linear $q^{+ a}$ multiplet, eq. \p{Freeq}. For the particular class of HK models, with
the Gibbons-Hawking \cite{GH} four-dimensional HK metrics as the bosonic target space, this form of the superfield off-shell action was
found in \cite{gpr}. Here we see that the same simple unique superfield action describes the most general 4n-dimensional HK ${\cal N}=4$ model.
The entire information about the (local) geometry of the underlying HK manifold is encoded in the off-shell harmonic constraint \p{coq-n},
with ${\cal L}^{+3a}$ as in \p{HK2}. The non-trivial structure of the component action comes out as a result
of imposing this nonlinear constraint. Note that the component actions in the HK case are also simplified: the full $\lambda$ frame Lagrangian
is given by the single term ${\cal L}^{(1)}$ defined in \p{Lagr1}, with ${\cal F}_{ab} = \Omega_{ab}$. In particular, there are
no quartic fermionic terms (the same is true of course for the $\tau$ frame Lagrangian \p{taulagr1}). The general HK ${\cal N}=4$ action \p{ActHK}
accompanied by the appropriate WZ coupling \p{WZcomp} provide the Lagrangian description of the generic HK ${\cal N}=4$ SQM models
discussed, in the Hamiltonian approach, in \cite{KLW}.

\subsection{HKT geometries conformal to HK ones}

One can choose some non-trivial Lagrangian ${\cal L}(f,u)$ leading to ${\cal F}_{[a, b]} \neq \Omega_{[a,b]}$, still keeping
the special form \p{HK2} for ${\cal L}^{+3a}(f^+, u)$. In this case, the vielbein coefficients, connections, as well as bridges from
the $\lambda$ basis and frames to the $\tau$ ones, are fully specified in terms of the HK potential ${\cal L}^{+ 4}$ by the same
relations as in the pure HK case. The basic differences from the latter case consist, first,  in the presence of the torsion \p{tor1} and, second,
in that the $\tau$ world metric \p{Metr} involves the coordinate-dependent ``symplectic'' metric $G_{[\und{c}\, \und{b}]}$.

All formulas are radically simplified in the $n=1$ case, that is for four-dimensional target spaces, where
\be
G_{[\und{c}\, \und{b}]} = \epsilon{}_{\und{c} \und{b}}\, G(f)\,, \quad G^{[\und{c}\, \und{b}]} = \epsilon{}^{\und{c} \und{b}}\, G^{-1}(f)\,.
\ee
The $\tau$ frame metric \p{Metr} then becomes
\be
g_{ia\;kb} = G\,\epsilon{}_{\und{c}\und{d}}\,\epsilon{}_{\,\und{l}\und{t}}\, e^{\,\und{l}\und{c}}_{\,ia}\, e^{\,\und{t}\und{d}}_{\,kb}
= G\,h_{ia\;kb}\,,
\ee
where $h_{ia\,kb} = \epsilon{}_{\und{c}\und{d}}\,\epsilon{}_{\,\und{l}\und{t}}\, e^{\,\und{l}\und{c}}_{\,ia}\,
e^{\,\und{t}\und{d}}_{\,kb}$ is the HK metric computed by the HK potential ${\cal L}^{+ 4}\,$. Since the bridge $M_{a}^{\und{b}}$ is
 a matrix in the fundamental representation of $USp(2) \sim SU(2)$, the connection $D^{\;\;\;\und{b}}_{\und{i}\,\und{a}\;\;\;\und{d}} $
 is traceless,
\be
D^{\;\;\;\und{b}}_{\und{i}\,\und{a}\;\;\;\und{b}} = 0\,,\lb{trVanish}
\ee
whence
\be
\nabla_{\und{k}\und{b}} G_{[\und{c}\,\und{d}]} = \epsilon{}_{\und{c}\und{d}}\left(e^{kb}_{\und{k}\und{b}}\,\partial_{kb}\, G +
D_{\und{k}\und{b}\;\;\;\und{e}}^{\;\;\;\und{e}}\,G\right) = \epsilon{}_{\und{c}\und{d}} \,e^{kb}_{\und{k}\und{b}}\,\partial_{kb}\, G\,,
\ee
and
\be
C_{\und{i}\und{a}\;\und{k}\und{b}\;\und{l}\und{c}} = \epsilon_{\und{i}\und{l}}\epsilon_{\und{c}\und{b}}\,
e_{\und{k}\und{a}}^{k a}\, \partial_{ka}\,G
+ \epsilon_{\und{k}\und{l}}\epsilon_{\und{a}\und{c}}\,\,e_{\und{i}\und{b}}^{ib}\,\partial_{ib}\, G\,. \lb{tor22}
\ee

Thus in this case the metric is conformal to the HK one, with the conformal factor $G(f)$. The torsion is also simply expressed
through $G(f)$.

In the general case, with unconstrained $G(f)$, the torsion is not closed, so this geometry is a particular case of weak HKT
geometry. The strong HKT arises under the constraint \p{close2} which, in the four-dimensional case, can be checked to be equivalent
to
\be
\Delta\,G = h^{ia\,kb }\,\nabla_{ia}\partial_{kb}\,G = 0\,. \lb{BoxCond}
\ee
Here,
\be
\nabla_{ia}\partial_{kb} = \partial_{ia}\partial_{kb} - \Gamma(h)^{\;ld}_{ia\,kb}\partial_{ld}\,,
\ee
with $\Gamma(h)^{\;ld}_{ia\,kb}$ being the standard Christoffel symbol for the HK metric $h_{ia\,kb}$, so that
$\Delta$ is the covariant Laplace-Beltrami operator on the given 4-dimensional HK manifold. While bringing \p{close2}
to the form \p{BoxCond}, we made use of the
condition \p{trVanish} which, in virtue of the representation \p{Iden5}, amounts to the following constraints on the $\tau$ frame vielbeins:
\be
e^{kb}_{\und{t}\und{a}}\,e^{lc}_{\und{l}\und{b}}\left(\epsilon^{\und{l}\und{t}}\,\partial_{[kb}e^{\und{j}\und{b}}_{lc]} +
 \epsilon^{\und{j}\und{t}}\,\partial_{[kb}e^{\und{l}\und{b}}_{lc]}\right) = 0\,.
\ee

In the general $n>1$ case $G_{\und{a}\,\und{b}}\neq
\Omega_{\und{a}\,\und{b}}G$ and the metric \p{Metr} is not conformal
to the corresponding $4n$-dimensional HK one, though the vielbeins
and bridges are the same as in the HK geometry.  The differential
operator in the constraint \p{close2} which yields the
$4n$-dimensional strong HKT geometry in this case, is some
generalization of the standard Beltrami-Laplace operator. Note that
for $n>1$ the conformally flat ansatz
$G_{\und{a}\,\und{b}}=\Omega_{\und{a}\,\und{b}}G$ gives rise by the
cyclic identity (\ref{Cycle3}) to $G=const$ and, hence, to the
torsionless HK geometry\begin{footnote}{E.I. thanks Andrei Smilga
for a discussion on this point.}\end{footnote}.  If
$G_{\und{a}\,\und{b}}\neq \Omega_{\und{a}\,\und{b}}G$, the torsion
is given by the general formula \p{tor1}.

{}From the above consideration we conclude that  one is able to
construct some HKT metric with torsion from any HK metric. The
relevant geometry is weak HKT in the generic case, but it becomes
strong HKT after imposing the appropriate constraint on the
conformal factor (in the 4-dimensional case), or some natural matrix
generalization of this factor (for $4n$-dimensional targets).

It should be stressed that this method of generating HKT geometries from the HK ones is well known
for four-dimensional targets (the so-called Callan-Harvey-Strominger ansatz) \cite{chs,GiPaSt}. In the present setting, it is recovered as
a subclass of HKT geometries associated with the off-shell $d=1$ supermultiplets $\bf{(4,4,0)}$. There naturally arises a
generalization of this ansatz to higher-dimensional target manifolds.

\subsection{HKT geometries with general ${\cal L}^{+ 3 a}$}

We first consider the simplified case with the ``free'' choice \p{constHK} for the potential ${\cal L}(f^\pm, u)$, but
with an arbitrary analytic potential ${\cal L}^{+ 3 a}(f^+, u)$. Now, because
\be
{\cal F}_{[a\, b]} = \Omega_{[a \,b]} = const\,, \lb{Restr}
\ee
the part $\hat{C}_{ia\;kb\;lc}$ of the torsion defined in \p{tor3} and \p{defChat}  vanishes,
$$
\hat{C}_{ia\;kb\;lc} = 0\,,
$$
and the torsion is given by
\be
C_{ia\;kb\;lc} = \partial_{ia}\,{\cal B}_{kb\;lc} + \partial_{lc}\,{\cal B}_{ia\;kb} + \partial_{kb}\,{\cal B}_{lc\;ia}\,, \lb{tor33}
\ee
with
\be
{\cal B}_{ia\;kb} = -\int du \,\Omega_{[g\,f]}\left(E^{+ g}_{ia}E^{-f}_{kb} -  E^{+ g}_{kb}E^{-f}_{ia}\right). \lb{pot11}
\ee
Thus the torsion is closed and the corresponding geometry is strong HKT. Actually, this is just the geometry of the $(4, 0), d=2$
sigma models, and the analytic unconstrained potential ${\cal L}^{+ 3 a}(f^+, u)$ coincides with that introduced in \cite{dks} to solve
this geometry in the harmonic space approach. Note that the torsion potential \p{pot11} becomes pure gauge in the HK case
(to check this, one must use \p{HKnonan}), and the torsion \p{tor33} vanishes, as should be. Also note that
the vanishing of $\hat{C}_{ia\;kb\;lc}$
and the identical fulfillment of the general closedness equation \p{closed1} is actually achieved
under the condition
\be
\partial_{-d}{\cal F}_{[ a\,b]}= 0\,,
\ee
which is weaker than \p{Restr}: it implies ${\cal F}_{[ a\,b]}$ to be analytic. However, using the analytic
$\lambda$ frame $GL(2n)$ gauge freedom, one can come back to \p{Restr} as a gauge condition reducing this $GL(2n)$ gauge group
to its $USp(2n)$ subgroup defined in \p{gltosp}, \p{spnparam}.

The general $d=1$ superfield action for the HKT case under consideration takes the same simple form \p{ActHK} as in the pure HK case.

If there is a non-trivial scalar potential ${\cal L}(f, u)$, one has $\partial_{-a}{\cal F}_{[a b]} \neq 0$. As a result,
the term $\hat{C}_{ia\;kb\;lc}$ defined in \p{defChat} is non-vanishing and so it makes an extra contribution to the torsion.
The corresponding geometry is weak HKT, but it becomes strong HKT under the conditions derived in Sect. 5.6. Once again,
these conditions are simplified in the case of four-dimensional target manifolds. Namely, for generic ${\cal L}^{+ 3a}\,$,
an analog of the covariant harmonicity condition \p{BoxCond} is as follows
\be
\Delta\, G + \epsilon^{\und{i}\und{k}}\,\epsilon^{\und{a}\und{b}}\left({\cal V}_{\und{i}\,\und{a}}\,\nabla_{\und{k}\,\und{b}}  +
\nabla_{\und{i}\,\und{a}}{\cal V}_{\und{k}\,\und{b}}
+ {\cal V}_{\und{i}\,\und{a}}\,{\cal V}_{\und{k}\,\und{b}}\right)G = 0\,. \lb{BoxCond2}
\ee
Here,
\be
 {\cal V}_{\und{i}\,\und{a}} := D_{\und{i}\und{a}\;\;\;\und{b}}^{\;\;\;\und{b}}
\ee
and $\Delta $ is the covariant Laplace-Beltrami operator for the metric corresponding to the HKT geometry with
${\cal F}_{[ab]} = \Omega_{[ab]}$ and
the same ${\cal L}^{+ 3 a}\,$. Eq. \p{BoxCond2} resembles the condition of closedness of the torsion in the four-dimensional
strong HKT system with the metric admitting a tri-holomorphic isometry and with an extra three-vector field \cite{opf}. The
connection ${\cal V}_{\und{i}\,\und{a}}$ looks like a four-dimensional analog of this vector field.

In the gauge $G = const$ (which is the particular $USp(2)\sim SU(2)$ case of the $\tau$ gauge \p{TangGauge})
the constraint \p{BoxCond2} is simplified to the following one
\be
\left(\nabla^{\und{i}\,\und{a}} +  {\cal V}^{\und{i}\,\und{a}}\right) {\cal V}_{\und{i}\,\und{a}}= 0\,. \lb{BoxCond3}
\ee
It is instructive to be convinced that this equation is again a constraint on the $\lambda$ frame symplectic metric
${\cal F}_{[a\,b]}(f, u) =
{\cal F}(f, u)\,\epsilon_{a\,b}$. To this end, recall the relations \p{Redund} and \p{Gauge5} which in the present case,
up to the residual $USp(2)$ $\tau$ gauge freedom, amount to the following ones
\be
L^{\und{b}}_{\und{a}} = \delta^{\und b}_{\und{a}}\, {\tilde{G}}^{\,\frac{1}{2}}\,, \quad M^{\und{b}}_c =
{\tilde{G}}^{\,\frac{1}{2}}\,\tilde{M}^{\und{b}}_c\,, \qquad \tilde{G} =
\frac{1}{2}\int du\,\epsilon^{\und{a}\und{b}}\,\epsilon_{c d}\,(\tilde{M}^{-1})_{\und{b}}^c (\tilde{M}^{-1})_{\und{a}}^d\,{\cal F}(f, u)\,.
\ee
Further, one should take into account that the vector connection ${\cal V}_{\und{i}\und{a}}$ corresponding to the bridge
$M^{\und{b}}_c $ and appearing in \p{BoxCond3} is related to the connection $\tilde{{\cal V}}_{\und{i}\und{a}}$ associated with the ``minimal''
bridge $\tilde{M}^{\und{b}}_c $ via the following gauge transformation
\be
{\cal V}_{\und{i}\und{a}} = {\tilde{G}}^{-\frac{1}{2}}\left(\tilde{{\cal V}}_{\und{i}\und{a}}
+ \frac{1}{2}\,{\tilde{G}}^{-1}\tilde{\nabla}_{\und{i}\und{a}}\,\tilde{G} \right).
\ee
Substituting this into eq. \p{BoxCond3}, we reduce it just to the form \p{BoxCond2} with ``tildas'' on all involved quantities.
This equation can be treated as the alternative $\tau$ gauge choice $M^{\und{b}}_c = \tilde{M}^{\und{b}}_c$ in \p{BoxCond2}.

As follows from our consideration here, from any strong HKT metric with some ${\cal L}^{+3a}$,
one can construct new metrics of similar kind, both the weak and strong HKT ones. This observation
is true for any dimension of the HKT target space and is an obvious generalization of the relation between HKT and HK metrics discussed
in Subsect. 6.2.

\setcounter{equation}{0}
\section{${\cal N}=4$ supersymmetry in terms of ${\cal N}=2$ superfields}
As we have seen above, the general off-shell ${\cal N}=4$ supersymmetric Lagrangian for a set
of harmonic analytic superfields $q^{+ a}$ with the field content $({\bf 4, 4, 0})$ exhibits a weak HKT geometry for bosonic fields.
On the other hand, it is known that the most general target geometry implied by ${\cal N}=4$ supersymmetry
for $d=1$ sigma models with such multiplets is even weaker. In particular, the corresponding triplet of complex structures does not form
a quaternionic algebra and the covariant constancy condition for it is replaced by some weaker condition. The only possibility
to encompass this general situation within our approach is to include into the game, together with the
$({\bf 4, 4, 0})$ multiplets represented by the analytic superfields $q^{+ a}$,
also the so-called mirror $({\bf 4, 4, 0})$ multiplets \cite{IKL1,ivnie}. These multiplets have a different assignment
of their fields with respect to
the full $R$-symmetry group $SU(2)\times SU(2)$ of the ${\cal N}=4, d=1$ Poincar\'e supersymmetry and, accordingly,
different transformation properties under the latter\footnote{As observed in a recent paper \cite{Gat}, these different
$({\bf 4, 4, 0})$ multiplets are recovered by dimensional reduction from different ${\cal N}=(4,4), d=2$ twisted chiral multiplets.}.
In order to describe these mirror multiplets in a manifestly ${\cal N}=4$ supersymmetric
way on equal footing with those represented by $q^{+ a}$, one needs to resort to the bi-harmonic ${\cal N}=4$ superspace
which includes two different analytic subspaces associated with two independent sets of the harmonic variables for two $R$ symmetry $SU(2)$
groups \cite{ivnie}. The mutually mirror $({\bf 4, 4, 0})$ multiplets ``live'' as superfields on these two non-equivalent
${\cal N}=4, d=1$ harmonic analytic subspaces. Leaving the general analysis within such a bi-harmonic superfield framework for the future,
here we tackle the same problem in the ${\cal N}=2$ superfield formalism, thus continuing the consideration in Sect. 2.

As a first step, we start from the ${\cal N}=2$ superspace action \p{n2act} in Sect.2.2 and look for the conditions
which allow for one more supersymmetry
\begin{equation}
\delta Z^\alpha=\epsilon{\cal J}^\alpha_{\bar\beta}\bar D\bar Z^{\bar\beta},\quad
\delta\bar Z^{\bar\alpha}=\epsilon{\cal J}^{\bar\alpha}_{\beta} D Z^{\beta}, \lb{3super}
\end{equation}
where $\epsilon$ is a real Grassmann parameter. The indices $\alpha, \bar\beta$ take values $1, \ldots 2n$, i.e.
in the simplest $n=1$ case we deal with a doublet of chiral ${\cal N}=2$ superfields, so that the real dimension of the
corresponding bosonic target space is just 4. The conditions for the existence of extra supersymmetry \p{3super}
have been studied in \cite{Hu},
and we first recall the results obtained there. Preservation of chirality leads to the constraints
\begin{equation}
\partial_{[\alpha}{\cal J}^{\bar\gamma}_{\beta]}=0,\quad
\partial_{[\bar\alpha}{\cal J}^{\gamma}_{\bar\beta]}=0\,.\label{chirp}
\end{equation}
The supersymmetry algebra requires $\cal J$ to be a complex structure
\begin{equation}
{\cal J}^{\bar\gamma}_{\beta}{\cal J}^{\alpha}_{\bar\gamma}=-\delta^{\alpha}_{\beta} \label{comps}
\end{equation}
which satisfies the integrability conditions
\begin{equation}
\partial_\gamma({\cal J}^\alpha_{[\bar\beta}){\cal J}^\gamma_{\bar\delta]}=0,\quad
\partial_{[\bar\gamma}({\cal J}^\alpha_{\bar\beta]}){\cal J}^{\bar\gamma}_\tau=0,
\end{equation}
These two equations are in fact consequences of (\ref{chirp}) and (\ref{comps}). The conditions for the action (\ref{n2act})
to be invariant under this new supersymmetry are the hermiticity of the metric
\begin{equation}
{\cal J}^{\bar\gamma}_{(\beta}g_{\alpha)\bar\gamma}=0,\quad g_{\gamma(\bar\alpha}{\cal J}^{\gamma}_{\bar\beta)}=0\,,
\label{hermi2}
\end{equation}
together with the equation
\begin{equation}
\frac{1}{2}\partial_\alpha({\cal J}^\delta_{\bar\beta}g_{\delta\bar\gamma})
-\partial_\delta(g_{\alpha[\bar\beta}){\cal J}^\delta_{\bar\gamma]}
+\partial_\alpha(g_{\delta[\bar\beta}){\cal J}^\delta_{\bar\gamma]}
-\frac{1}{4}\partial_{[\bar\beta}(B_{\bar\gamma\bar\tau]}){\cal J}^{\bar\tau}_{\alpha}=0,
\label{symme}
\end{equation}
and its complex conjugate. In the ${\cal N}=2$ superspace framework these equations correspond
to the equations (\ref{symcov}) for the new complex structure, which state that the symmetrized covariant derivative
of the complex structure vanishes,
provided one of the symmetrized indices is barred and the other is unbarred. The equations (\ref{symcov}) for ${\cal J}_\beta^{\bar\delta}$,
when both symmetrized indices are of the same type, are automatically satisfied. Indeed, using (\ref{chirp}), (\ref{comps})
and the hermiticity (\ref{hermi2}) of the metric, one gets
\begin{equation}
\nabla_\alpha{\cal J}_\beta^{\bar\delta}=3g^{\gamma\bar\delta}\partial_{[\alpha}({\cal J}^{\bar\lambda}_\beta g_{\gamma]\bar\lambda}).
\end{equation}
Finally, one gets, as the last equation to ensure the invariance of the action, the following one:
\begin{equation}
\partial_{\bar\delta}(\partial_{[\alpha}(B_{\beta\gamma})){\cal J}^{\bar\delta}_{\tau]} =0, \label{diffi2}
\end{equation}
and its complex conjugate. These equations are a part of equations (\ref{diffi}) for the new complex structure. The object
appearing in eq. (\ref{diffi}) for ${\cal J}$ is a 4-form. Using the first complex structure $I$, this 4-form may be decomposed
into $(p, 4-p)$ forms. Equation (\ref{diffi2}) and its complex conjugate are the $(4,0)$ and $(0,4)$
parts of the equation (\ref{diffi}) for ${\cal J}$. It may be checked that the remaining parts of this equation
are automatically satisfied as a consequence of (\ref{torco}) (written for ${\cal J}$), (\ref{chirp}) and (\ref{symme}).

Let us suppose for a moment that we have ${\cal N}=4$ supersymmetry, and that the last supersymmetry transformation
is determined by some complex structure ${\cal J}'$. Moreover, we suppose that the three complex structures
form a quaternionic algebra. Then the third complex structure is the product of the first two, that is to say
\begin{equation}
{{\cal J}'}_{\bar\alpha}^\beta=i{\cal J}_{\bar\alpha}^\beta,\quad {{\cal J}'}_{\alpha}^{\bar\beta}=-i{\cal J}_{\alpha}^{\bar\beta}.
\end{equation}
When we apply equation (\ref{symme}) to the complex structures ${\cal J}$ and ${\cal J}'$, we immediately get the
following two equations
\begin{equation}
\frac{1}{2}\partial_\alpha({\cal J}^\delta_{\bar\beta}g_{\delta\bar\gamma})
-\partial_\delta(g_{\alpha[\bar\beta}){\cal J}^\delta_{\bar\gamma]}
+\partial_\alpha(g_{\delta[\bar\beta}){\cal J}^\delta_{\bar\gamma]}=0\,,\quad
\partial_{[\bar\beta}(B_{\bar\gamma\bar\tau]}){\cal J}^{\bar\tau}_{\alpha}=0\,.
\label{asymme}
\end{equation}
One easily deduces from these equations that the covariant derivatives of the complex structures vanish.
One thus recovers the statement  made in \cite{Strom}, that ${\cal N}=4, d=1$ supersymmetry with
complex structures satisfying the quaternionic algebra correspond to the weak HKT geometry, which differs from
the strong HKT geometry appearing in $d=2$ $(4,0)$ supersymmetric sigma models merely
in that the torsion 3-form is not closed.

The above consideration was based upon the assumption that the extra supersymmetries associated with the complex structures
${\cal J}$ and ${\cal J}'$ are uniformly realized on all involved chiral superfields $Z^\alpha$, that is by the transformation
law \p{3super} and by a similar law with ${\cal J}'$. This situation can be shown to be in one-to-one correspondence with
the analysis in the previous Sections, the $({\bf 4, 4, 0})$ multiplets being formed by pairs of the chiral ${\cal N}=2$
superfields $Z^\alpha$. Let us now turn to the most general situation, when, along with the standard $({\bf 4, 4, 0})$ multiplets, their
mirror counterparts are also taken into account. This option was not considered  in the literature before.

The simultaneous description of two different sorts  of the  $({\bf 4, 4, 0})$ multiplets in the ${\cal N}=2$ superspace
amounts to considering two types
of complex coordinates $z^\alpha$, $\alpha=1\cdots m$, and $u^a$, $a=1\cdots n$. The manifold under consideration has dimension $2(m+n)$.
For reasons to be clear soon, both $m$ and $n$ must be even. Both $z^\alpha$ and $u^a$ are assumed to be the lowest components
of the chiral ${\cal N}=2$ superfields $Z^\alpha$ and $U^a\,$.
Only the standard ${\cal N}=2$ supersymmetry is manifest, and we suppose that these superfields transform under an extra ${\cal N}=2$
supersymmetry as :
\begin{eqnarray}&&
\delta Z^\alpha=\epsilon J^\alpha_{\bar\beta}\bar D\bar Z^{\bar\beta},\quad \delta U^a=\bar \epsilon K^a_{\bar b}\bar D \bar{U}^{\bar b},\cr &&
\delta \bar Z^{\bar\alpha}=\bar\epsilon \bar J^{\bar \alpha}_{\beta} DZ^\beta,\quad \delta \bar U^{\bar a}
=\epsilon \bar K^{\bar a}_{b} DU^{b},\label{extsusy}
\end{eqnarray}
where $\epsilon$ is a new complex Grassmann parameter. Clearly, when considered together, these two sets of superfields possess
essentially different transformation laws under the extra supersymmetry.
Further, we shall make the hypothesis that the transformation of one type of coordinates depend only on the same type of complex coordinate:
\begin{equation}
\partial_a J^\alpha_{\bar\beta}=\partial_{\bar a}J^\alpha_{\bar\beta}=0,\quad
\partial_\alpha K^a_{\bar b}=\partial_{\bar\alpha} K^a_{\bar b}=0,
\end{equation}
(the same conditions are assumed for $\bar J$ and $\bar K$)\footnote{This hypothesis is substantiated by the fact already mentioned
that in the ${\cal N}=4$ bi-harmonic superfield approach the mutually mirror $({\bf 4, 4, 0})$ multiplets are represented by analytic
superfields  with different sorts of harmonic analyticity, so ${\cal N}=4$ supersymmetry cannot mix them.}.
The preservation of chirality imposes the constraints
\begin{equation}
\partial_{[\bar\alpha}J^\gamma_{\bar\beta]}=0,\quad \partial_{[\bar a}K^c_{\bar b]}=0,\label{chiralp}
\end{equation}
and analogous conditions for $\bar J$ and $\bar K$. The supersymmetry algebra requires
\begin{equation}
J^\alpha_{\bar\gamma}\bar J^{\bar\gamma}_\beta=-\delta^\alpha_\gamma,\quad
K^a_{\bar c}\bar K^{\bar c}_b=-\delta^a_b, \label{square}
\end{equation}
as well as
\begin{equation}
\partial_\gamma(J^\alpha_{[\bar\beta})J^\gamma_{\bar\delta]}=0,\quad
\partial_{\bar\gamma}(J^\alpha_{\bar\beta})\bar J^{\bar\gamma}_\tau
+J^\alpha_{\bar\gamma}\partial_{\bar\beta}(\bar J^{\bar\gamma}_\tau)=0,
\label{nieje}
\end{equation}
together with analogous equations for $K$ and $\bar K$. These last equations are consequences of (\ref{chiralp}) and (\ref{square}).
Let us combine $Z^\alpha,\bar Z^{\bar\alpha}$ and $U^a, \bar U^{\bar a}$ into a generic coordinate
$X^A$, with $A= \alpha, \bar\alpha, a, \bar a$.
By separating the Grassmann parameter into its real and imaginary part $\epsilon=\epsilon_1+i\epsilon_2$, we see that
we have introduced two new complex structures
\begin{equation} \delta X^A=\epsilon_1{\cal J}_1{\cal D}X^A+\epsilon_2{\cal J}_2{\cal D}X^A, \end{equation}
with ${\cal D}=D+\bar D$. These complex structures have the block-diagonal form:
\begin{equation}
{\cal J}_1=\left(\begin{array}{cccc}0&J&0&0\cr\bar J&0&0&0\cr 0&0&0&K\cr 0&0&\bar K&0\end{array}\right),\quad
{\cal J}_2=\left(\begin{array}{cccc}0&iJ&0&0\cr -i\bar J&0&0&0\cr 0&0&0& -iK\cr 0&0&i\bar K&0\end{array}\right),
\end{equation}
while the complex structure corresponding to the manifest ${\cal N}=2$ supersymmetry reads
\begin{equation}
{\cal J}_3=\left(\begin{array}{cccc}i\mathbf{1}&0&0&0\cr 0&-i\mathbf{1}&0&0\cr 0&0& i\mathbf{1}&0 \cr 0&0&0& -i\mathbf{1}
\end{array}\right).
\end{equation}
It is to be noticed that these three complex structures do not form a quaternionic algebra:
\begin{equation}
{\cal J}_1{\cal J}_2\neq{\cal J}_3,\quad {\cal J}_2{\cal J}_3\neq{\cal J}_1, \lb{nonquater}
\end{equation}
although they have definite anticommutation relations
\begin{equation}\{ {\cal J}_a,{\cal J}_b\} =2\delta_{ab}\mathbf{1}. \lb{anticomm}
\end{equation}

The general ${\cal N}=2$ superfield action relevant to our consideration is as follows\footnote{It can be cast in the generic form
(\ref{n2act}) by combining chiral superfields $Z^\alpha$ and $U^a$ and their conjugates into the new sets $\tilde{Z}^A = (Z^\alpha, U^a)$
and $\bar{\tilde{Z}}^{\bar{A}} = (\bar Z^{\bar\alpha}, \bar U^{\bar a})$.}:
\begin{eqnarray}&&
S=\frac{1}{2}\int dtd\theta d\bar\theta\left(-(g_{\alpha\bar\beta}DZ^\alpha\bar D\bar Z^{\bar\beta}
+g_{\alpha\bar b}DZ^\alpha\bar D\bar U^{\bar b}+g_{a\bar\beta}DU^a\bar D\bar Z^{\bar\beta}+
g_{a\bar b}DU^a\bar D\bar U^{\bar b}) \right.\cr && +\frac{1}{12}(B_{\alpha\beta}DZ^\alpha DZ^\beta +
B_{\alpha b}DZ^\alpha DU^b +B_{a \beta}DU^a DZ^\beta +B_{ab}DU^a DU^b) \cr &&\left.
+\frac{1}{12}(B_{\bar\alpha\bar\beta}\bar D\bar Z^{\bar\alpha} \bar D\bar Z^{\bar\beta} +
B_{\bar\alpha \bar b}\bar D\bar Z^{\bar\alpha}\bar D\bar U^{\bar b}
+B_{\bar a\bar \beta}\bar D\bar U^{\bar a}\bar D\bar Z^{\bar\beta}
+B_{\bar a\bar b}\bar D\bar U^{\bar a}\bar D\bar U^{\bar b})\right).
\end{eqnarray}
Invariance of this action under the transformations (\ref{extsusy}) leads to the equations
\begin{equation}
g_{\alpha\bar b}=0,\quad g_{\alpha(\bar\beta}J^{\alpha}_{\bar\gamma)}=0, \quad g_{a(\bar b}K^{a}_{\bar c)}=0,
\end{equation}
together with their complex conjugates. This set of equations may be interpreted as coming from the hermiticity
of the metric with respect to the complex structures ${\cal J}_2$ and ${\cal J}_3$. One furthermore obtains the conditions
\begin{eqnarray}&&
\partial_\alpha(g_{\delta\bar\gamma}J^\delta_{\bar\beta})
-2\,\partial_\alpha(g_{\delta[\bar\gamma})J^\delta_{\bar\beta]}
+2\,\partial_\delta(g_{\alpha[\bar\gamma})J^\delta_{\bar\beta]} =0\,,\cr &&
\partial_a(g_{d\bar c}K^d_{\bar b})
-2\,\partial_a(g_{d[\bar c})K^d_{\bar b]}
+2\,\partial_d(g_{a[\bar c})K^d_{\bar b]} = 0\,,\cr &&
[2\,\partial_{[\bar\beta}(B_{\bar\gamma]\bar d})+ \partial_{\bar d}(B_{\bar\beta\bar\gamma})]\bar K^{\bar d}_a
+ 6\, \partial_a(g_{\lambda[\bar\beta})J^\lambda_{\bar\gamma]} =0,\cr &&
[2\,\partial_{[\bar b}(B_{\bar c]\bar\delta}) + \partial_{\bar\delta}(B_{\bar b\bar c})]
\bar J^{\bar\delta}_\alpha + 6\,\partial_\alpha(g_{e[\bar b})K^e_{\bar c]}=0,\cr &&
\partial_{[\alpha}B_{\gamma\delta]}=0,\quad \partial_{[a}B_{cd]}=0,
\end{eqnarray}
together with the complex conjugate equations. When taking into account the previous constraints, these equations
can be cast in the same form as the equation (\ref{symcov}), but in application to the new complex structures ${\cal J}_1$ and ${\cal J}_2$.
In the condensed notation through the coordinates $X^A$, these equations imply that the symmetrized covariant derivatives
of these complex structures vanish
\begin{equation}
\nabla_{(A}{{\cal J}_a}^C_{B)}=0\,, \quad a=1,2, \lb{eqJ12}
\end{equation}
but they do not imply that the {\it full} covariant derivatives of the complex structures vanish. The properties \p{eqJ12}, \p{nonquater}
and \p{anticomm} are just the characteristic properties of the complex structures in the most general target geometry of the ${\cal N}=4, d=1$
supersymmetric sigma models based on multiplets $({\bf 4, 4, 0)}$ \cite{GiPaSt,Hu}.

Finally, using the notation
\begin{eqnarray}
c_{\alpha\beta c}=\frac{1}{3}(\partial_cB_{\alpha\beta}+\partial_\alpha B_{\beta c}+\partial_\beta B_{c\alpha}),&&\cr
c_{ab\gamma}=\frac{1}{3}(\partial_\gamma B_{ab}+\partial_a B_{b\gamma}+\partial_b B_{\gamma a}),&&
\end{eqnarray}
one gets, as the last conditions for the invariance of the action under ${\cal N}=4$ supersymmetry, the following relations
\begin{eqnarray}
\partial_{\bar e}(c_{\alpha[bc})\bar K^{\bar e}_{d]}=0,
\quad \partial_{\bar e}(c_{\alpha\beta[c})\bar K^{\bar e}_{d]}=0, &&\cr
\partial_{\bar \tau}(c_{a[\beta\gamma})\bar J^{\bar \tau}_{\delta]}=0,
\quad \partial_{\bar \tau}(c_{ab[\gamma})\bar J^{\bar \tau}_{\delta]}=0, &&\label{cric}
\end{eqnarray}
together with their complex conjugates. As stated above, these constraints coincide with a particular sector
of the geometric equations (\ref{diffi}), this time applied to the complex structures ${\cal J}_1$ and ${\cal J}_2$.

To summarize, we have shown that the most general bosonic target geometry of ${\cal N}=4\,$, $d=1$ sigma models can be achieved
provided both types of the mutually mirror $({\bf 4, 4, 0})$ multiplets are simultaneously included. Keeping in mind that $m=2k$ and $n= 2l$,
$k, l \geq 1$, such a generalized geometry can be realized only for the target spaces of the dimension $d = 4(k +l) = 4s\,, s = 2, 3\ldots\,$,
with the minimal dimension $d=8\,$. The ${\cal N}=4, d=1$ sigma models built only on the multiplets of the same sort exhibit
weak HKT geometry as their target geometries.

\setcounter{equation}{0}

\section{Summary and conclusions}
In this paper, we constructed and studied the sigma-model and Wess-Zumino-type superfield actions for any number of $({\bf 4, 4, 0})$ multiplets
described by the analytic harmonic superfields $q^{+ a}$ subjected to the most general harmonic constraints
which are compatible with the $d=1$ harmonic analyticity. The main distinguishing feature of our superfield description as compared, e.g.,
with the formulations used in \cite{CoPa,GiPaSt,Hu}, is that our actions exhibit manifest off-shell ${\cal N}=4$ supersymmetry. We have shown that the
most general bosonic target space geometry of this class of $d=1$ sigma model actions is the weak HKT geometry. Our superfield
approach suggests that this geometry is solved in terms of two unconstrained potentials: the general non-analytic scalar
potential ${\cal L}(f^{+ a}, f^{- a}, u^{\pm i})$ and the analytic potential ${\cal L}^{+ 3a}(f^{+ b}, u^{\pm i})\,$. We presented
the general expressions for the relevant metric and torsion, as well as the conditions under which the weak HKT geometry becomes the strong HKT and
the ordinary HK ones. We also presented the general component form of the analytic WZ term which involves the coupling
of the multiplets $({\bf 4, 4, 0})$ to an external abelian self-dual gauge field given on a target HKT manifold.

Furthermore, using the ${\cal N}=2$ superfield formalism, we have shown that the most general ${\cal N}=4, d=1$
sigma-model target geometry arises, when taking into account simultaneously two types of $({\bf 4, 4, 0})$ multiplets which are mirror
with respect to each other. Actually, our consideration in this paper together with the results of refs. \cite{ils} and \cite{ivnie} give a hint as to
what could be independent primary potentials solving this general geometry. Indeed, as was already mentioned, the ordinary
and mirror $({\bf 4, 4, 0})$ multiplets are represented by superfields living on two distinct analytic subspaces of the bi-harmonic ${\cal N}=4$
superspace $(t, \theta^{i\alpha}, u^{\pm i}, v^{\pm \alpha})$ where $i$ and $\alpha$ are doublet indices of two commuting automorphism ($R$-symmetry)
$SU(2)$ groups of the ${\cal N}=4, d=1$ Poincar\'e supergroup (see \cite{ivnie} for more details). Then the general sigma model superfield
action should be an obvious generalization of the action \p{nonlqAct}, with both types of the analytic superfields and their harmonic derivatives
involved. Also, there should be imposed appropriate harmonic constraints of the type \p{coq-n}, giving rise to some extended set of potentials
of the type ${\cal L}^{+3 a}$, each depending on the superfields of one or another harmonic analyticity. This set of bi-harmonic analytic
potentials together with the Lagrangian depending simultaneously on both types of $({\bf 4, 4, 0})$ superfields can be thought of as constituting
the minimal set of the underlying potentials of the general ${\cal N}=4, d=1$ sigma-model target geometry. Note that the general action of one
ordinary and one mirror {\it linear} $({\bf 4,4,0})$ multiplets (i.e. described by analytic bi-harmonic superfields with harmonic constraints
of the type \p{coq}) was constructed in \cite{ils} using the ordinary ${\cal N}=4$ superfields. It was shown that this action
possesses ${\cal N}=8$ supersymmetry under some simple restrictions on the superfield Lagrangian\footnote{The pair of mutually mirror
$({\bf 4, 4, 0})$ multiplets form an irreducible multiplet $({\bf 8, 8, 0})$ of ${\cal N}=8, d=1$ supersymmetry \cite{BIKL}, while
this  is not true for pairs of the $({\bf 4, 4, 0})$ multiplets of the same type.}. It is of obvious interest to extend our consideration here
to the bi-harmonic stuff, and, in particular, to find general conditions of existence of ${\cal N}=8$ supersymmetry in the system of linear
and nonlinear ordinary and mirror $({\bf 4, 4, 0})$ multiplets. The dimension of the target spaces in such ${\cal N}=8$ supersymmetric models
is a multiple of 8, and the corresponding geometries should reveal a close relation to octonions \cite{GiPaSt,Top8}.

Besides extension to the bi-harmonic approach, there are a few other directions in which the results presented here can be applied
and advanced. For instance, using our general geometric consideration, it would be interesting to construct couplings
to external non-abelian gauge fields on the general HKT manifolds and their various particular cases. The basic ingredients of such
a construction in the case of linear ${\cal N}=4, d=1$ multiplets \cite{ikons} are the ``semi-dynamical''  spin
$({\bf 4, 4, 0})$ multiplets \cite{spin} described solely by the superfield WZ terms of the kind \p{WZq}. It is natural to try
to extend this construction to the case, when both the coordinate and spin  multiplets are general non-linear $({\bf 4, 4, 0})$
multiplets introduced in the present paper. Also, it is tempting to apply, to the general case, the superfield  procedure
of gauging isometries \cite{gpr,DI2,DI3}, which produces the off-shell actions of the multiplets $({\bf n, 4, 4-n})\,$, ${\bf n}=0,1,2,3\,$,
from those of the $({\bf 4, 4, 0})$ multiplets. In this way, most general ${\cal N}=4$ supersymmetric sigma-model and WZ-type actions
of such multiplets can be set up. An interesting task for future study is to quantize the ${\cal N}=4$ mechanics models constructed here,
with taking into account both the sigma-model and WZ-type couplings, and to establish relations between these quantum models and various
geometric objects of the target geometry, like Dirac and Dolbeault complexes and their possible generalizations, along the lines
of refs. \cite{KLW} and \cite{ism}.

\setcounter{equation}{0}
\section*{Acknowledgments}
The authors thank Maxim Konyushikhin and Andrei Smilga for interest in the work and valuable discussions. The work of E.I. was supported
in part by RFBR grants 09-02-01209, 09-01-93107, 09-02-91349 and a grant of the Heisenberg-Landau program.
He thanks Laboratoire de Physique, ENS-Lyon, for the kind hospitality multiply extended to him in the course of this
work. The final stage of this research was completed, while E.I. was visiting SUBATECH, University of Nantes, under
the Convention N${}^{\rm o}$ 2010 11780. He is indebted to the Directorate of SUBATECH for the kind hospitality.

\end{document}